    \RequirePackage{fix-cm}
    \documentclass[smallextended]{svjour3}       
    \smartqed  
    \usepackage{amsmath,latexsym,amssymb,amsfonts,epsfig}
    \usepackage[usenames, dvipsnames]{color}
    \usepackage[english]{babel}
    \usepackage[utf8]{inputenc}			
    \usepackage{graphicx, texdraw}   
    \usepackage{bm}

    \usepackage{lmodern}
    \usepackage{mathtools}
    \usepackage{mathrsfs} 
    
    \definecolor{OliveGreen}{rgb}{0,0.6,0}
    \definecolor{darkGreen}{rgb}{0,0.4,0}
    \definecolor{darkBlue}{rgb}{0.,0.,0.6}

    \usepackage[left= 2.1cm, right = 2.1cm, top= 3cm, bottom = 3cm]{geometry}     

    \usepackage{enumerate}

    \usepackage[colorlinks=true,citecolor=darkBlue, linkcolor=blue]{hyperref}

    \usepackage{csquotes}

    \usepackage{amsmath}
    \usepackage{amsfonts}
    \usepackage{amssymb}
    \usepackage{braket}

    \usepackage[numbers,sort&compress]{natbib}

    \usepackage[normalem]{ulem} 

    \usepackage[misc]{ifsym} 

\newcommand{\vect}[1]{\boldsymbol{#1}}

\newcommand{\boldNabla}{\boldsymbol{\nabla}}

\newcommand{\kS}{\kappa_\mathrm{S}}

\newcommand{\kB}{\kappa_\mathrm{B}}

\newcommand{\alphaB}{\alpha_\mathrm{B}}

\newcommand{\betaB}{\beta_\mathrm{B}}

\newcommand{\R}{\vect{r}}
\newcommand{\Intd}{\mathrm{d }}

\newcommand{\eR}{\vect{e}_r}
\newcommand{\ePhi}{\vect{e}_{\phi}}

\newcommand{\eZ}{\vect{e}_{z}}

\newcommand{\gOne}{\vect{g}_1}
\newcommand{\gTwo}{\vect{g}_2}

\newcommand{\JS}{J_{\mathrm{S}}}

\newcommand{\vRot}{\vect{v}^{\mathrm{R}}}
\newcommand{\pRot}{p^{\mathrm{R}}}
\newcommand{\vRotcom}{v^{\mathrm{R}}}

\newcommand{\vIm}{\vect{v}^{*}}
\newcommand{\pIm}{p^{*}}
\newcommand{\vImcom}{v^{*}}

\newcommand{\EB}{E_\mathrm{B}}

\newcommand{\uu}{\vect{u}}
\newcommand{\vv}{\vect{v}}

\newcommand{\tehta}{\theta}

\newcommand{\phanSp}{\phantom{cccc}}
\newcommand{\phanSpSp}{\phantom{ccccccccccccc}}

\usepackage{csquotes}

\begin{document}

    \title{Slow rotation of a spherical particle inside an elastic tube
    }
    \subtitle{}


    \author{Abdallah Daddi-Moussa-Ider         \and
	    Maciej Lisicki   \and
	    Stephan Gekle
    }


    \institute{   \Letter~  A. Daddi-Moussa-Ider and S. Gekle \at
		  Biofluid Simulation and Modeling, Fachbereich Physik, Universit\"at Bayreuth, Universit\"{a}tsstra{\ss}e 30, Bayreuth 95440, Germany \\
		  \email{abdallah.daddi-moussa-ider@uni-bayreuth.de}           
	      \and
	      M. Lisicki  \at
		  Department of Applied Mathematics and Theoretical Physics, Wilberforce Rd, Cambridge CB3 0WA, United Kingdom \\
		  Institute of Theoretical Physics, Faculty of Physics, University of Warsaw, Pasteura 5, 02-093 Warsaw, Poland
    }

    \date{Received: April 27, 2017 / Accepted: date}

    \maketitle

    \begin{abstract}
    In this paper, we present an analytical calculation of the rotational mobility functions of a particle rotating on the centerline of an elastic cylindrical tube whose membrane exhibits resistance towards shearing and bending.
    We find that the correction to the particle rotational mobility about the cylinder axis depends solely on membrane shearing properties while both shearing and bending manifest themselves for the rotational mobility about an axis perpendicular to the cylinder axis. 
    In the {quasi-steady limit of vanishing frequency}, the particle rotational mobility nearby a no-slip rigid cylinder is recovered only if the membrane possesses a non-vanishing resistance towards shearing. 
    We further show that for the asymmetric rotation along the cylinder radial axis, a coupling between shearing and bending exists.
    Our analytical predictions are compared and validated with corresponding boundary integral simulations where a very good agreement is obtained.

    
    \end{abstract}

    \section{Introduction}

    The assessment of effects of geometric confinement on the motion of microparticles in a viscous fluid is of great importance, since such conditions are found in numerous biological or industrial processes~\cite{sharp87, hernandez05}. In such systems, the long-range hydrodynamic interactions which determine macroscopic transport coefficients, are significantly modified due to the flows reflected from the confining boundaries~\cite{happel12, cichocki88b, cichocki99, Dlugosz_2015}. Many of the works have been devoted to motion in tubular channels for their relevance to transport of fluids in microfluidic systems~\cite{Squires2005, wang13} or in human arteries~\cite{Frey1952}. Notably, an important property of these networks of channels is the elasticity of their building material. Blood flow in capillaries relies on the collagen and elastin filaments within their wall, which enable them to deform in response to changing pressure~\cite{Shadwick1999,Caro2011}. 

    Theoretical modeling of slow viscous dynamics and hydrodynamics of particles in narrow channels has been mostly focused on flows within hard cylindrical tubes.  The monograph of Happel and Brenner~\cite{happel12} encompasses most theoretical results available. Axial motion of a point particle has been studied extensively due to relevance to rheology measurements~\cite{faxen22, wakiya53, faxen59, bohlin60, greenstein67, greenstein68, sano87, zimmerman04}, with later extensions to account for the finite size~\cite{Leichtberg_1976} or non-spherical shape~\cite{Yeh_2013}. The motion perpendicular to the axis has been further studied by  Hasimoto~\cite{hasimoto76}.

    The first attempt to address the slow symmetric rotation of a sphere in an infinitely long hard cylinder dates back to Haberman~\cite{haberman61} and later by Brenner and Sonshine~\cite{brenner64} who gave the torque acting on the rotating sphere as power series of the ratio of particle to cylinder diameter.
    The rotation of an axisymmetric body within a circular cylinder of finite length has been investigated by Brenner~\cite{brenner64b} using the point couple approximation technique. 
    The frictional force~\cite{greenstein76} and torque~\cite{greenstein75} exerted on a slowly rotating eccentrically positioned sphere within an infinitely long circular cylinder has been studied by Greenstein and coworkers.
    The latter further investigated the slow rotation of two spheres placed about the cylinder axis in a direction perpendicular to their line of centers~\cite{greenstein70}.
    Complementary theoretical works have been conducted by Hirschfeld and coworkers~\cite{hirschfeld72, hirschfeld84} to determine the cylindrical wall effects on the translating-rotating particle of arbitrary shape.
    Additionally, perturbative solutions for the rotation of eccentric spheres flowing in a cylindrical tube have been derived by T\"{o}zeren~\cite{tozeren82, tozeren84, tozeren83}, finding a good agreement  with the previous solutions. 
    Modeling of hydrodynamic interactions involving a torus or a circular orifice~\cite{chen13} has been further presented~\cite{oneill96}.

    Despite an abundance of results available for hard confining boundaries, not many studies focus on the role of elasticity on the motion of microparticles in confinement. Observations of flow through a deformable elastic channel~\cite{Rubinow1972, fung13} demonstrate phenomena that can be related to the cardiovascular and respiratory systems, including the generation of instabilities~\cite{bertram89, shankar09, shankar10}, propagation of small-amplitude waves~\cite{grotberg01, grotberg04}, and hysteretic shearing of arterial walls~\cite{canic06}. 
    The flexibility in microfluidic devices has also been indicated as a potential way of controlling flow~\cite{stone04, holmes13}. More recent works have been devoted to the influence of elastic tube deformation on flow behavior of a shear-thinning fluid~\cite{nahar12, nahar13} or the steady flow in thick-walled flexible elastic tubes~\cite{mikelic07, marzo05}. 
    No theoretical studies, however, explore the role of elastic confinement on the hydrodynamic mobility of particles. 
    
    This motivates us to compute the flow field generated by a particle rotating inside a realistically modeled elastic channel. 
    We have modeled the membrane using the neo-Hookean model for shearing~\cite{ramanujan98, barthes11, lac04, barthes16}, and the Helfrich model~\cite{helfrich73, zhong89, Guckenberger_2017} for bending of its surface. An analogous approach has been successfully applied to the motion of small particles in the presence of planar membranes~\cite{felderhof06, felderhof06b, daddi16,daddi16c,daddi17}, between two elastic sheets~\cite{daddi16b} and in the vicinity of a spherical elastic capsule~\cite{daddi17b, daddi17c}. The theoretical results presented in some of these works have been favorably compared with fully resolved boundary integral method (BIM) simulations, and thus constitute a practical approximate tool for analysis of confined motion in elastically bounded systems.
    The present study computes the frequency-dependent rotational mobility corrections due to the elastic confinement which has not been previously analyzed.

    The remainder of the paper is organized as follows.
    In Sec.~\ref{theoretical}, we derive analytical expressions for the flow field induced by a point-torque oriented either parallel or perpendicular to the cylinder axis, by expressing the solutions of the Stokes equations in terms of Fourier-Bessel integrals. 
    We then compute in Sec.~\ref{mobilityAndDeformation} the leading order self- and pair-mobility functions for the rotation along or perpendicular to the cylinder axis. 
    Moreover, the membrane displacement field induced by the particle for a given actuation is presented.
    For a given set of parameters, we compare in Sec.~\ref{comparisonWithBIM} our analytical predictions with fully resolved boundary integral simulations, where a good agreement is obtained.
    Concluding remarks are offered in Sec.~\ref{conclusions}.
    The appendix outlines the main derivation steps for the determination of the linearized traction jumps stemming from membrane shearing and bending rigidities.


    \section{Theoretical description}\label{theoretical}

\begin{figure}
    \begin{center}
      \includegraphics[scale=1.2]{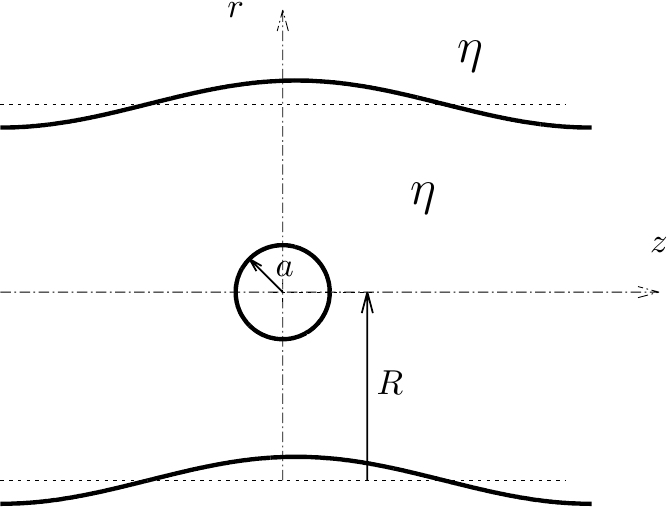}
    \caption{Illustration of the system setup. A small spherical particle of radius $a$ placed at the origin rotating nearby an elastic tube of undisturbed radius $R$.} 
    \label{cylinderIllustration}
    \end{center}
    \end{figure}

    We consider a small solid spherical particle of radius $a$, placed on the axis of a cylindrical elastic tube of undisturbed radius $R \gg a$.
    {The fluid inside and outside the tube is assumed to be incompressible of the same shear viscosity $\eta$.}
    {An oscillatory torque acts on the particle inducing periodic rotational motion whose amplitude is linearly related to the amplitude of the acting torque. Our final goal is to compute the rotational mobility representing the coefficient of proportionality between torque and motion.}
    We employ the cylindrical coordinate system $(r,\phi,z)$ where $r$ is the radius, $\phi$ is the azimuthal angle and $z$ is the axial direction along the cylinder axis with the origin located at the center of the particle (see Fig.~\ref{cylinderIllustration} for an illustration of the system setup). 
    The flow fields inside and outside the cylindrical channel are labeled 1 and 2, respectively.
    
    We proceed by computing the rotlet solution which follows from the solution of the forced Stokes equations 
    \begin{subequations}
    \begin{align}
    \eta \boldNabla^2 \vect{v}_1 - \boldNabla p_1 + \vect{F} (\R) &= 0 \, , \\
    \boldNabla \cdot \vect{v}_1 &= 0 \, , 
    \end{align}
    \label{equationMotion_fluid1}
    \end{subequations}
    inside the tube (for $r<R$) and 
    \begin{subequations}
    \begin{align}
    \eta \boldNabla^2 \vect{v}_2 - \boldNabla p_2 &= 0 \, , \\
    \boldNabla \cdot \vect{v}_2 &= 0 \, , 
    \end{align}
    \label{equationMotion_fluid2}
    \end{subequations}
    outside (for $r>R$). 
    {Here $\vect{F}(\R)$ represents an arbitrary time-dependent force density acting on the fluid. We specifically consider a distribution that has only the asymmetric dipolar term
    \begin{align}
    \oint_S \vect{r}\times \vect{F} \, \mathrm{d}S  = \vect{L} \, , 
    \end{align}
    where the integral is taken over the surface $S$ of the spherical particle. Since the particle is small, we shall consider its point-like limit. Then the antisymmetric dipolar term in the force multipole expansion yields the flow field induced by a rotlet of strength $\vect{L}$. The flow field around a rotlet in an infinite fluid is given by~\cite{blake71}
    \begin{equation}
    \vect{v}(\vect{r})= \frac{1}{8\pi\eta} \frac{\vect{L}\times\vect{r}}{r^3} \, .
    \end{equation}
    Our aim is to find the corresponding solution in the space confined by an elastic cylindrical tube.}
    
    For realistic parameters, we have shown in earlier work~\cite{daddi16} that the term with a time derivative in the unsteady Stokes equations leads to a negligible contribution to the total mobility corrections and thus is not considered in the present work.

    Eqs.~\eqref{equationMotion_fluid1} and \eqref{equationMotion_fluid2} are subject to the regularity conditions 
    \begin{align}
    |\vect{v}_1| < \infty \,  &\text{ for } |\R| = 0 \, , \label{regularity_Inside} \\
    \vect{v}_1 \to \mathbf{0} \, &\text{ for } z \to \infty \, , \label{regularity_Inside_2} \\
    \vect{v}_2 \to \mathbf{0} \, &\text{ for } |\R| \to \infty \, , \label{regularity_Outside}
    \end{align}
    together with the boundary conditions imposed at the undisplaced membrane $r=R$.    
       {This commonly used simplification is justified since we are dealing with small deformations only. 
       In other situations, when the finite amplitude of deformation is important, it becomes necessary to apply the boundary conditions at the displaced membrane, see for instance Refs.~\cite{sekimoto93, weekley06, salez15, saintyves16, rallabandi16}.}
        The velocity field across the membrane is continuous, leading to
    \begin{align}
      [v_r] &= 0 \, , \label{BC:v_r} \\
      [v_\phi] &= 0 \, , \label{BC:v_phi} \\
      [v_z] &= 0 \, , \label{BC:v_z}
    \end{align}
    while the elastic membrane introduces a discontinuity in the fluid stress tensor
    \begin{align}
      [\sigma_{\phi r}] &= \Delta f_\phi^{\mathrm{S}}  \, , \label{BC:sigma_r_phi} \\
      [\sigma_{zr}] &= \Delta f_z^{\mathrm{S}}  \, , \label{BC:sigma_r_z} \\
      [\sigma_{rr}] &= \Delta f_r^{\mathrm{S}} + \Delta f_r^{\mathrm{B}} \, , \label{BC:sigma_r_r}
    \end{align}
    with the notation $[w] := w(r = R^{+}) - w(r = R^{-})$ refering to the jump of a given quantity $w$ across the membrane. 
    The fluid stress tensor is expressed in cylindrical coordinates as~\cite{kim13}
    \begin{align}
      \sigma_{\phi r} &= \eta \left( v_{\phi ,r} - \frac{v_\phi + v_{r, \phi}}{r} \right) \, , \notag \\
      \sigma_{zr}     &= \eta (v_{z,r} + v_{r,z}) \, , \notag  \\
      \sigma_{rr}     &= -p + 2\eta v_{r,r} \, . \notag 
    \end{align}
    
    The traction jumps can be decomposed into a contribution due to the in-plane shearing elasticity (superscript S) and a contribution stemming from membrane bending rigidity (superscript B). 
    Shearing is accounted for using the neo-Hookean model~\cite{ramanujan98}.
    As derived in the Appendix, the linearized traction jumps due to shearing elasticity are written as
    \begin{subequations}
    \begin{align}
    \Delta f_\phi^{\mathrm{S}} &= -\frac{\kS}{3} \left( u_{\phi, zz} + \frac{3 u_{z, \phi z}}{R} + \frac{4 (u_{r,\phi} + u_{\phi, \phi\phi})}{R^2} \right)  \, , \\
    \Delta f_{z}^{\mathrm{S}}    &= -\frac{\kS}{3} \left( 4 u_{z,zz} + \frac{2 u_{r,z} + 3 u_{\phi, z\phi}}{R} + \frac{u_{z, \phi\phi}}{R^2} \right) \, , \\
    \Delta f_{r}^{\mathrm{S}}    &= \frac{2\kS}{3} \left( \frac{2 (u_r + u_{\phi, \phi}) }{R^2} + \frac{u_{z,z}}{R}  \right) \, ,
    \end{align}
    \end{subequations}
    where $\kS$ is the elastic shear modulus.
    The comma in indices denotes a partial spatial derivative. 

    Bending of the membrane is described following the Helfrich model~\cite{helfrich73, Guckenberger_2017} as
    \begin{equation}\label{bendingTraction}
      \Delta f_r^{\mathrm{B}} = \kB \Big( R^3 u_{r,zzzz} + 2R(u_{r,zz} + u_{r,zz\phi\phi}) 
		  + \frac{u_r+2u_{r,\phi\phi}+u_{r, \phi\phi\phi\phi}}{R} \Big)  \, ,
    \end{equation}
    where $\kB$ is the bending modulus. 
    Moreover, $\Delta f_\phi^{\mathrm{B}} = \Delta f_z^{\mathrm{B}} = 0 $ since bending does not introduce a discontinuity in the tangential traction jumps~\cite{Guckenberger_2017}.

{Similar as above, we apply the no-slip boundary condition at the undisplaced membrane surface \cite{bickel06}
    \begin{equation}
   \frac{\partial \uu (\phi, z)}{\partial t} = \vv (r, \phi, z)|_{r = R} \, ,
    \end{equation}
     which in Fourier space is written as 
    \begin{equation}
    \uu (\phi, z) = \left. \frac{\vv (r, \phi, z)}{i\omega} \right|_{r = R} \, . 
    \label{no-slip-relation}
    \end{equation}
 }

    Having introduced the regularity and boundary conditions, we then solve the equations of fluid motion by expanding them in the form of Fourier-Bessel integrals.
    For this aim, solutions will be searched for in the two distinct regions i.e. inside and outside the cylindrical membrane separately. 
    We write the solution in terms of integrals of harmonic functions with unknown coefficients, which we then determine from the boundary conditions.

    We begin by expressing the solution of Eqs.~\eqref{equationMotion_fluid1} inside the cylinder as a sum of a point-torque (point-couple) flow field and the flow field reflected from the membrane, as
    \begin{align}
    \vect{v}_1 &= \vRot + \vIm \, , \notag \\
    p_1        &= \pRot + \pIm \, , \notag
    \end{align}
    where $\vRot$ and $\pRot$ are the rotlet solution in an unbounded medium and $\vIm$ and $\pIm$ are the solutions of the homogenous Stokes equations
    \begin{subequations}
    \begin{align}
    \eta \boldNabla^2 \vIm - \boldNabla \pIm &= 0 \, , \\
    \boldNabla \cdot \vIm &= 0 \, ,
    \end{align}
    \label{equationMotion_fluid1_Image}
    \end{subequations}
    required to satisfy the regularity and boundary conditions.
    In the next section, we shall first consider the axisymmetric rotational motion about the cylinder axis.

    \subsection{Axial rotlet}

    The solution for a point-torque of strength $\vect{L}=L_z \vect{e}_z$, located at the origin and directed along the $z$ direction reads~\cite{blake71}
    \begin{equation}
    \vRotcom_x = -\frac{L_z}{8\pi\eta} \frac{y}{d^3} \, , \qquad 
    \vRotcom_y = \frac{L_z}{8\pi\eta} \frac{x}{d^3} \, , \qquad
    \vRotcom_z = 0 \, ,  \notag
    \end{equation}
    and $ \pRot = 0$.
    Here $d:=\sqrt{r^2+z^2}$ is the distance from the rotlet position.
    Therefore, the velocity field is purely directed along the azimuthal direction such that
    \begin{equation}
    \vRotcom_r = 0 \, , \qquad  \vRotcom_\phi = \frac{L_z}{8\pi\eta} \frac{r}{d^3} = -\frac{L_z}{8\pi\eta} \frac{\partial}{\partial r} \frac{1}{d} \, . \label{azimuthalVelocity_Axial}
    \end{equation}

    By making use of the integral relation~\cite{watson95, brenner58}
    \begin{equation}
    \frac{1}{d} = \frac{2}{\pi} \int_0^\infty K_0(q r) \cos q z \, \Intd q \, ,  \label{firstFormula}
    \end{equation} 
    wherein $K_0$ is the zeroth order modified Bessel function of the second kind~\cite{abramowitz72}, the integral representation of the azimuthal fluid velocity field due to a point-torque reads
    \begin{equation}
    \vRotcom_\phi = \frac{L_z}{4\pi^2 \eta} \int_0^\infty q K_1(qr) \cos qz \, \Intd q \, . \label{v_phi_rotlet_Axial}
    \end{equation}

    For symmetric rotation about the cylinder axis, the homogenous Stokes equations Eqs.~\eqref{equationMotion_fluid2} and \eqref{equationMotion_fluid1_Image} reduce to,  
    \begin{equation}
      v_{\phi, rr}^* + \frac{v_{\phi,r}^*}{r}- \frac{v_\phi^*}{r^2} +v_{\phi, zz}^* = 0 \, ,
    \end{equation}
    and analogously for ${v_\phi}_2$.
    Using the method of separation of variables \cite{haberman83}, and by making use of the regularity equations stated by Eqs.~\eqref{regularity_Inside} through \eqref{regularity_Outside}, the image solution and external fluid velocity can therefore be presented in integral form as~\cite{brenner64} 
    \begin{subequations}
    \begin{align}
    v_\phi^*   &=  \frac{L_z}{4\pi^2\eta} \int_0^\infty A^*(q) I_1(qr) \cos qz \, \Intd q \, , \label{v_phi_star_Axial} \\
    {v_\phi}_2 &=  \frac{L_z}{4\pi^2\eta} \int_0^\infty A_2(q) K_1(qr) \cos qz \, \Intd q \, . \label{v_phi_2_Axial}
    \end{align}
    \end{subequations}

    The azimuthal velocity component across the membrane is continuous in virtue of Eq.~\eqref{BC:v_phi} leading to
    \begin{equation}
      K_1 A_2 - I_1 A^* = \frac{sK_1}{R} \, , \label{Eq_1_Axial}
    \end{equation}
    where $s:=qR$. 
    The modified Bessel functions have the argument $s$ which is dropped here for brevity.
    The unknown functions $A^*$ and $A_2$ are to be determined from the imposed traction jumps at the membrane.

    The discontinuity of the azimuthal-normal component of the fluid stress jump stated by Eq.~\eqref{BC:sigma_r_phi} leads to
    \begin{equation}
      (sI_0-I_1)A^* + \left(  \left(1-\frac{i\alpha s^2}{2} \right)K_1 + s K_0 \right) A_2 
      = \frac{s(sK_0+K_1)}{R} \, , \label{Eq_2_Axial}
    \end{equation}
    {where we have defined the shearing coefficient as
    \begin{equation}
    \alpha:=\frac{2 \kS}{3\eta R\omega}\, ,
    \end{equation}
    which quantifies the effect of shearing for a given actuation frequency $\omega$.}

    Solving Eqs.~\eqref{Eq_1_Axial} and \eqref{Eq_2_Axial} for the unknown coefficients $A^*$ and $A_2$ we obtain
    \begin{align}
    A^* &= \frac{1}{R} \frac{i\alpha s^2 K_1^2}{(2I_0-i\alpha s I_1)K_1 + 2K_0 I_1} \, , \label{A_star} \\
    A_2 &= \frac{1}{R} \frac{2s (I_0K_1 + I_1 K_0)}{(2K_0-i\alpha s K_1)I_1 + 2I_0 K_1} \, . \label{A_2}
    \end{align}

    {Interestingly, the coefficients $A^*$ and $A_2$ and thus the inner and outer flow fields depend solely on shear and do not depend on bending.}
    In particular, for $\alpha = 0$, the image solution Eq.~\eqref{v_phi_star_Axial} vanishes and the solution outside the cylinder \eqref{v_phi_2_Axial} is identical to the rotlet solution given by Eq.~\eqref{v_phi_rotlet_Axial}.

    In the limit $\alpha\to\infty$ {corresponding to the quasi-steady limit} of vanishing actuation frequency, or equivalently to an infinite membrane shearing modulus, we recover the result obtained earlier by Brenner~\cite{brenner64b}, namely
    \begin{equation}
    \lim_{\alpha\to\infty} A^* = -\frac{s K_1}{R I_1} \, , \notag
    \end{equation}
    and $A_2 = 0$ for which the outer fluid is stagnant.
    In the following, the solution for a radial rotlet will be derived.


    \subsection{Radial rotlet}

    Without loss of generality, we shall assume that the rotlet is exerted along the $x$ direction.
    The induced velocity field reads~\cite{blake71}
    \begin{equation}
    \vRotcom_x = 0 \, , \qquad
    \vRotcom_y = -\frac{L_x}{8\pi\eta} \frac{z}{d^3} \, , \qquad
    \vRotcom_z = \frac{L_x}{8\pi\eta} \frac{y}{d^3} \, , \notag
    \end{equation}
    and $\pRot = 0$.
    Transforming to cylindrical coordinates, we obtain
    \begin{equation}
    \vRotcom_r = -\frac{L_x}{8\pi\eta} \frac{z \sin \phi}{d^3} \, , \quad
    \vRotcom_\phi = -\frac{L_x}{8\pi\eta} \frac{z \cos \phi}{d^3} \, , \quad
    \vRotcom_z = \frac{L_x}{8\pi\eta} \frac{r \sin \phi}{d^3} \, . \notag
    \end{equation}

    After making use of Eq.~\eqref{firstFormula} together with~\cite{watson95, brenner58}
    \begin{equation}
    \frac{z}{d} = \frac{2}{\pi} \,  r \int_0^\infty K_1(q r) \sin q z \, \Intd q \, , \label{secondFormula}
    \end{equation}
    and by noting that
    \begin{equation}
    \frac{z}{d^3} = - \frac{1}{r} \frac{\partial}{\partial r} \frac{z}{d} \, , \qquad \frac{r}{d^3} = -\frac{\partial}{\partial r} \frac{1}{d}  \, , \label{notableRelation}
    \end{equation}
    the rotlet solution can therefore be expressed in an integral form as
    \begin{subequations}
     \begin{align}
    \vRotcom_r &= -\frac{L_x}{4\pi^2\eta} \, \sin \phi \int_0^\infty q K_0(qr) \sin qz \, \Intd q \, , \\
    \vRotcom_\phi &= -\frac{L_x}{4\pi^2\eta} \, \cos \phi \int_0^\infty q K_0(qr) \sin qz \, \Intd q \, , \\
    \vRotcom_z &= \frac{L_x}{4\pi^2\eta} \, \sin \phi \int_0^\infty q K_1(qr) \cos qz \, \Intd q \, .
    \end{align}
    \end{subequations}

    The reflected flow can be represented by using the fact that the homogenous Stokes equations~\eqref{equationMotion_fluid1_Image} have a general solution expressed in terms of 
    three harmonic functions $\Phi$, $\Psi$ and $\Gamma$ as~\cite[p. 77]{happel12}
    \begin{subequations}\label{vstar_radial}
    \begin{align}
    \vImcom_r &= {\Psi}_{,r} + \frac{{\Gamma}_{,\phi}}{r} + r \, {\Phi}_{,rr} \, , \label{v_r_prime_radial} \\
    \vImcom_\phi &= \frac{{\Psi}_{,\phi}}{r} - {\Gamma}_{,r} - \frac{{\Phi}_{,\phi}}{r} + {\Phi}_{,\phi r} \, , \label{v_phi_prime_radial} \\ 
    \vImcom_z &= {\Psi}_{,z} + r \, {\Phi}_{,rz} + {\Phi}_{,z} \, , \label{v_z_prime_radial} \\
    \pIm   &= -2\eta \, {\Phi}_{,zz} \, . \label{pressure_prime_radial}
    \end{align}
    \end{subequations}
    The functions $\Psi$, $\Phi$ and $\Gamma$ are solutions to the Laplace equation which can be written in an integral form as
    \begin{subequations}
     \begin{align}
    \Phi &= \frac{L_x}{4\pi^2\eta} \, \sin \phi \int_0^\infty {\varphi} (q) {g} (qr) \sin qz \, \Intd q \, , \\
    \Psi &= \frac{L_x}{4\pi^2\eta} \, \sin \phi \int_0^\infty \psi(q) {g} (qr) \sin qz \, \Intd q \, , \\
    \Gamma &= \frac{L_x}{4\pi^2\eta} \, \cos \phi \int_0^\infty \gamma(q) {g} (qr) \sin qz \, \Intd q \, ,
    \end{align}
    \end{subequations}
    {where $\varphi$, $\psi$ and $\gamma$ are wavenumber-dependent unknown functions to be determined from the underlying boundary conditions.}
    {Moreover, $g$ is a solution of the first order modified Bessel equation \cite{abramowitz72}. Since the solution needs to be regular at the origin owing to Eq.~\eqref{regularity_Inside}, we take $g \equiv I_1$ for the image solution,}    
    directly leading to
    \begin{subequations}
     \begin{align}
    v_r^*    &= \frac{L_x}{4\pi^2\eta} \frac{\sin \phi}{r} \int_0^\infty \bigg(\left( (2+q^2 r^2)I_1(qr)-qr I_0(qr) \right) {\varphi}^*(q) + \left(qrI_0(qr)-I_1(qr)\right)\psi^*(q) \notag \\
	      &\phanSpSp  - I_1(qr) \, \gamma^*(q) \bigg) \sin qz \, \Intd q \, , \\
    v_\phi^* &= \frac{L_x}{4\pi^2\eta} \frac{\cos \phi}{r} \int_0^\infty \bigg( \left( qr I_0(qr)-2I_1(qr) \right) {\varphi}^*(q) + I_1(qr)\psi^*(q) \notag \\ 
	      &\phanSpSp + \left( I_1(qr)-qrI_0(qr) \right)\gamma^*(q) \bigg) \sin qz \, \Intd q \, , \label{v_phi_image} \\
    v_z^*    &= \frac{L_x}{4\pi^2\eta} \sin \phi \int_0^\infty q \left( qr I_0(qr) {\varphi}^*(q) + I_1(qr) \psi^*(q)\right) \cos qz \, \Intd q  \, . \label{v_z_image} \\
    p^*      &= \frac{L_x}{2\pi^2} \sin \phi \int_0^\infty q^2 I_1(qr) {\varphi}^*(q) \sin qz \, \Intd q \, .
    \end{align}
    \end{subequations}
    {Since the solution has to decay at infinity in virtue of Eq.~\eqref{regularity_Outside}, we thus take $g \equiv K_1$ for the fluid outside, leading to}
    \begin{subequations}
     \begin{align}
    {v_r}_2    &= \frac{L_x}{4\pi^2\eta} \frac{\sin \phi}{r} \int_0^\infty \bigg( \left( (2+q^2 r^2)K_1(qr)+qr K_0(qr) \right) {\varphi}_2(q) -\left(qrK_0(qr)+K_1(qr)\right)\psi^*(q)  \notag \\
		&\phanSpSp - K_1(qr) \, \gamma^*(q) \bigg) \sin qz \, \Intd q \, , \\
    {v_\phi}_2 &= \frac{L_x}{4\pi^2\eta} \frac{\cos\phi}{r} \int_0^\infty \bigg( -\left(qr K_0(qr)+2K_1(qr)  \right) {\varphi}_2(q) + K_1(qr)\psi^*(q)  \notag \\
		&\phanSpSp + \left( K_1(qr)+qrK_0(qr) \right)\gamma^*(q) \bigg) \sin qz \, \Intd q \, , \\
    {v_z}_2    &= \frac{L_x \sin \phi}{4\pi^2\eta}  \int_0^\infty q \left( -qr K_0(qr){\varphi}_2(q) + K_1(qr) \psi^*(q) \right) \cos qz \, \Intd q  \, , \\
    p_2        &= \frac{L_x \sin \phi }{2\pi^2} \int_0^\infty q^2 K_1(qr) {\varphi}_2 (q) \sin qz \, \Intd q \, .
    \end{align}
    \end{subequations}
    The continuity of the fluid velocity field across the membrane as stated by Eqs.~\eqref{BC:v_r} through \eqref{BC:v_z} leads to
    \begin{align}
    \left( sI_0-(2+s^2)I_1\right){\varphi^*} + (I_1-sI_0)\psi^*+I_1\gamma^* -(K_1+sK_0)\psi_2
    + \left( sK_0+(2+s^2)K_1 \right) {\varphi_2}-K_1\gamma_2 &= -s K_0  \, , \label{KOntin_1} \\
    (2I_1-sI_0) {\varphi^*}-I_1 \psi^*+(sI_0-I_1)\gamma^*-(sK_0+2K_1) {\varphi_2} + K_1 \psi_2 
    +(K_1+sK_0) \, \gamma_2 &= -sK_0 \, , \\
    -s^2 I_0 {\varphi^*} -s I_1 \psi^*-s^2 K_0 {\varphi_2}+sK_1 \psi_2 &= s K_1 \, . \label{KOntin_3}
    \end{align}

    The unknown functions ${\varphi}_2$, $\psi_2$ and $\gamma_2$ associated to the external flow field can readily be expressed in terms of ${\varphi}^*$, $\psi^*$ and $\gamma^*$ {by solving Eqs.~\eqref{KOntin_1} through \eqref{KOntin_3} to obtain}
    \begin{align}
    {\varphi}_2   &= \frac{S {\varphi}^*+(K_1+sK_0)G \psi^* - K_1 G \gamma^*}{D} \, , \label{Eq_1} \\
    \psi_2   &= \frac{s \left( (2+s^2)K_0+s K_1 \right)G {\varphi}^*+S \psi^* - s K_0 G \gamma^*}{D} + 1 \, , \label{Eq_2} \\
    \gamma_2 &= \frac{2sK_0 G {\varphi}^* + 2K_1 G \psi^* }{D} 
	     + \frac{\left( S-G\left( sK_0+(2+s^2)K_1 \right) \right) \gamma^*}{D}  - 1 \, , \label{Eq_3}
    \end{align}
    where we have defined
    \begin{align}
    S &= -sK_0K_1 \left( sI_0+(2+s^2)I_1 \right) 
	    - s^2 \left( sI_0 K_0^2+I_1K_1^2 \right) \, , \notag \\
    G &= -s \left( I_0 K_1 + I_1 K_0 \right) \, , \notag  \\
    D &= s \big( s^2 K_0^3 + sK_0^2 K_1 - sK_1^3 
	    - (2+s^2)K_0 K_1^2 \big) \, . \notag
    \end{align}

    The expressions of ${\varphi}^*$, $\psi^*$ and $\gamma^*$ may be determined given the membrane constitutive model. 
    In the following, explicit analytical expressions will be derived by considering independently an idealized membrane with pure shearing or pure bending.

    \subsubsection{Pure shearing}
    
    As a first model, we consider an idealized elastic membrane with pure shearing resistance, such as an artificial capsule~\cite{rao94, bacher17}.
    The traction jump along the azimuthal direction given by Eq.~\eqref{BC:sigma_r_phi} depends only on membrane shearing resistance.
    We obtain
    \begin{equation}
    \begin{split}
    &\left( (4+s^2) I_1 -2sI_0 \right){\varphi}^* + (sI_0-2I_1)\psi^* + \left( sI_0 - (2+s^2)I_1 \right) \gamma^* +  \big( \left( i\alpha (8+3s^2)-(4+s^2) \right)K_1 \\
    &\phanSp+ 2s \left( i\alpha (2+s^2) - 1 \right) K_0 \big) {\varphi}_2 + \frac{1}{2} \left( \left( 4+2s^2 - i\alpha \left(8+s^2 \right) \right) K_1 +  s \left( 2 - i\alpha \left(4+s^2 \right) \right) K_0 \right) \gamma_2 \\
    &\phanSp\phanSp+ \left( 2\left(1-i\alpha (2+s^2)\right) K_1 + s(1-2i\alpha) K_0 \right) \psi_2 
    = -s^2 K_1 \, .
    \end{split} \label{Eq_4_RR}
    \end{equation}
    The traction jump along the axial direction stated by Eq.~\eqref{BC:sigma_r_z} is also independent of bending leading to
    \begin{equation}
    \begin{split}
    & s^2(I_0+sI_1) {\varphi}^* + s(sI_0 - I_1) \psi^* + s \left( s \left( 1+i\alpha(3+2s^2) \right)K_0  +  \left( i\alpha(5+s^2)-s^2 \right) K_1 \right) {\varphi}_2 \\
    &\phanSp+ s \left( \left( 1-i\alpha (3+2s^2) \right) K_1  + s(1-i\alpha) K_0 \right) \psi_2 - \frac{i\alpha s}{2} \left( 3sK_0 + 5K_1 \right) \gamma_2 = -s(sK_0+K_1) \, .
    \end{split} \label{Eq_5_RR}
    \end{equation}
    By considering only the shearing contribution in the normal traction jump in Eq.~\eqref{BC:sigma_r_r} we get
    \begin{equation}
      2s^2 I_1 {\varphi}^* + \left( i\alpha s(4+s^2)K_0 + 2 \left( i\alpha (4+s^2)-s^2 \right) K_1 \right) {\varphi}_2 - i\alpha \left( 2s K_0 + (4+s^2) K_1 \right) \psi_2 
      - 2i\alpha (sK_0 + 2K_1) \gamma_2 = 0 \, . \label{Eq_6_RR}
    \end{equation}

    Eqs.~\eqref{Eq_1} through \eqref{Eq_6_RR} form a closed system of equations for the unknown functions.
    Due to their complexity, analytical expressions are not listed here.
    In particular, in the limit $\alpha \to \infty$ we obtain
    \begin{align}
    \lim_{\alpha\to\infty} {\varphi}^*   &= \frac{(I_0K_1+I_1K_0)(2I_1-sI_0)}{s(sI_0-I_1)(I_0^2-I_1^2) - 2I_0 I_1^2} \, , \label{phi_S_hard} \\
    \lim_{\alpha\to\infty} \psi^*   &= \frac{sI_0^2(sK_0-K_1)+I_0 I_1(s^2K_1-2sK_0+2K_1)-s I_1^2 K_1}{s(sI_0-I_1)(I_0^2-I_1^2) - 2I_0 I_1^2} \, , \label{psi_S_hard} \\
    \lim_{\alpha\to\infty} \gamma^* &= \frac{(s^2K_0+sK_1+4K_0)I_1^2+2I_0I_1K_1-sI_0^2(sK_0+K_1)}{s(sI_0-I_1)(I_0^2-I_1^2) - 2I_0 I_1^2} \, , \label{omega_S_hard}
    \end{align}
    where the functions ${\varphi}_2$, $\psi_2$ and $\gamma_2$ vanish in this limit.

    \subsubsection{Pure bending}
    
    Another membrane model involves only a bending resistance, as commonly considered for fluid vesicles~\cite{bukman96, luo13}.
    Neglecting the shearing contribution in the traction jump along the $\phi$ direction, Eq.~\eqref{BC:sigma_r_phi} reads
    \begin{equation}
    \begin{split}
    &\left( (4+s^2) I_1 -2sI_0 \right){\varphi}^* + (sI_0-2I_1)\psi^* + \left( sI_0 - (2+s^2)I_1 \right) \gamma^*  - \big(  (4+s^2) K_1 + 2s K_0 \big) {\varphi}_2 + \left( 2 K_1 + s K_0 \right) \psi_2 \\
    &\phanSp+ \left( \left( 2+s^2  \right) K_1 + s K_0 \right) \gamma_2 = -s^2 K_1 \, .
    \end{split} \label{Eq_4_RR_Bending}
    \end{equation}
    The traction jump across the $z$ direction in the absence of shearing is continuous leading to
    \begin{equation}
    \begin{split}
     s(I_0+sI_1) {\varphi}^* + (sI_0 - I_1) \psi^* + s \left( K_0 -s K_1 \right) {\varphi}_2 +  \left( K_1 +  s K_0 \right) \psi_2  = -(sK_0+K_1) \, .
    \end{split} \label{Eq_5_RR_Bending}
    \end{equation}
    while the normal traction jump is discontinuous leading to
    \begin{equation}
    \begin{split}
      &2I_1 {\varphi}^* +\left( i\alphaB^3 s^2 \left( sK_0+(2+s^2)K_1 \right) - 2K_1 \right) {\varphi}_2 -i\alphaB^3 s^2 (sK_0+K_1) \psi_2 - i\alphaB^3 s^2 K_1 \gamma_2 = 0 \, , \label{Eq_6_RR_Bending}
    \end{split}
    \end{equation}
    {where we have defined the bending coefficient $\alphaB$ as
    \begin{equation}
      \alphaB := \frac{1}{R} \left(\frac{\kB}{\eta\omega}\right)^{1/3} \, , 
    \end{equation}
    quantifying the effect of bending.}

    By plugging the expressions of ${\varphi}_2$, $\psi_2$ and $\gamma_2$ as given by Eqs.~\eqref{Eq_1} through \eqref{Eq_3} into Eqs.~\eqref{Eq_4_RR_Bending} through \eqref{Eq_6_RR_Bending}, expressions for ${\varphi}^*$, $\psi^*$ and $\gamma^*$ can be obtained.
    In particular, by taking the limit $\alphaB \to \infty$ the coefficients read
    \begin{align}
    \lim_{\alphaB\to\infty} {\varphi}^* &= \frac{-sK_0(sK_0+K_1)}{sK_0 \left(2sI_0 - (3+s^2)I_1 \right) + (3+s^2)(sI_0 - 2I_1) K_1} \, , \notag \\
    \lim_{\alphaB\to\infty} \psi^* &= \frac{sK_0 \left( sK_0 + (2+s^2)K_1 \right)}{sK_0 \left(2sI_0 - (3+s^2)I_1 \right) + (3+s^2)(sI_0 - 2I_1) K_1} \, , \notag \\
    \lim_{\alphaB\to\infty} \gamma^* &= \frac{2s K_0 K_1}{sK_0 \left(2sI_0 - (3+s^2)I_1 \right) + (3+s^2)(sI_0 - 2I_1) K_1} \, , \notag 
    \end{align}
    which are in contrast to the solution for a hard-cylinder with stick boundary conditions given by Eqs.~\eqref{phi_S_hard} through \eqref{omega_S_hard}.
    This difference is explained by the fact that bending following the Helfrich model does not lead to a discontinuity in the tangential traction jumps~\cite{Guckenberger_2017}.
    Moreover, the normal traction jump as stated by Eq.~\eqref{bendingTraction} depends uniquely on the radial (normal) displacement and does not involve the tangential displacements $u_\phi$ and $u_z$.
    As a result, even by taking an infinite membrane bending modulus, the tangential displacements are still completely free.
    This behavior therefore cannot represent the rigid cylinder limit where membrane deformation in all directions must be restricted.  
     Such behavior has previously been observed near spherical membranes as well~\cite{daddi17b, daddi17c}.

    \subsubsection{Shearing and bending}

    For a membrane endowed simultaneously with shearing and bending rigidities, a similar resolution procedure can be employed.
    Explicit analytical expressions can be obtained via computer algebra systems, but they are rather complicated and are therefore not listed here.
    We further mention that a coupling between shearing and bending exists, meaning that the solutions derived above for pure shearing and bending cannot be added up linearly. 
    This coupling behavior has previously been observed for two parallel planar~\cite{daddi16b} or spherical membranes~\cite{daddi17b, daddi17c}, in contrast to the single membrane case where adding up linearly the shearing and bending related solutions holds~\cite{daddi16, daddi17}.

    {In order to clarify the mentioned coupling between shear and bending, consider two different idealized membranes, one with pure bending resistance ($\alpha=0$) and another one with pure shear resistance ($\alphaB=0$).} 
{For a membrane endowed simultaneously with both shear and bending rigidities, we have shown in Eqs.~\eqref{Eq_1}--\eqref{Eq_3} that the unknown functions outside the tube $X_2$ are related to the functions inside $X^*$ in the following way
\begin{equation}
 X_2 = A X^* + b \, , \label{exactSystem}
\end{equation}
where $X_2 = (\varphi_2, \psi_2, \gamma_2)^\mathrm{T}$, $X^* = (\varphi^*, \psi^*, \gamma^*)^\mathrm{T}$, $A$ is a $3\times 3$ known matrix and $b= (0,1,-1)^\mathrm{T}$. }

{We now denote by ${X_2}_\mathrm{S}$, ${X_2}_\mathrm{B}$ the solutions outside the tube for a membrane with pure shear and pure bending, respectively, and by ${X}_\mathrm{S}^*$, ${X}_\mathrm{B}^*$ the corresponding image system solutions.
Accordingly, 
\begin{equation}
 {X_2}_\mathrm{S} = A X_\mathrm{S}^* + b \, , \qquad 
 {X_2}_\mathrm{B} = A X_\mathrm{B}^* + b \, , 
\end{equation}
leading after taking the sum member by member to
\begin{equation}
 \hat{X}_2 = A \hat{X}^* + 2b \, , \label{superpositionEqn}
\end{equation}
where $\hat{X}_2 = {X_2}_\mathrm{S} + {X_2}_\mathrm{B} $ and $\hat{X}^* = X_\mathrm{S}^* + X_\mathrm{B}^*$ are the superposition solutions.
Clearly, this relation is different from the original equation~\eqref{exactSystem} since $b \ne 0$, and therefore the true solutions $X_2$ and $X^*$ cannot both be identical to the superposed functions $\hat{X}_2$ and $\hat{X}^*$.
As a consequence, they cannot satisfy the correct boundary conditions showing that shear and bending are coupled and cannot be added up linearly.
}

    \section{Particle rotational mobility and membrane deformation}\label{mobilityAndDeformation}

    The rotlet solution obtained in the previous section serves as a basis for the determination of the particle rotational mobilities along and perpendicular to the cylinder axis.
    We restrict our present consideration to the point-particle approximation, and thus the particle size is much smaller than the cylinder radius.
    We shall show that this approximation, despite its simplicity, can lead to a surprisingly good agreement with boundary integral simulations of truly extended particles.

    \subsection{Axial rotational mobility}

    Beginning with the rotational motion symmetrically around the cylinder axis, the leading-order correction to the rotational mobility of a point-particle is
    \begin{equation}
    \Delta \mu_{\parallel}^{\mathrm{S}} = L_z^{-1}  \lim_{\R\to 0} {\Omega_z^*} \, , \label{Delta_mu_zz_def}
    \end{equation}
    with 
    \begin{equation}
    {\Omega_z^*} = \frac{1}{2} \left( v_{\phi,r}^* + \frac{v_\phi^*}{r} \right) \, , \notag
    \end{equation}
    being the $z$ component of the correction to the fluid angular velocity ${\boldsymbol{\Omega}^*} := \tfrac{1}{2} \boldNabla \times \vect{v}^*$.  Making use of Eq.~\eqref{v_phi_star_Axial}, we obtain
    \begin{equation}
    \Delta \mu_{\parallel}^{\mathrm{S}} = \frac{1}{8\pi^2\eta} \int_0^\infty q A^* \, \Intd q \, . \notag 
    \end{equation}
    Scaling by the bulk rotational mobility $\mu_0^{rr} = 1/(8\pi\eta a^3)$, 
    the scaled frequency-dependent correction to the rotational mobility takes the form
    \begin{equation}
    \frac{\Delta \mu_{\parallel}^{\mathrm{S}}}{\mu_0^{rr}} = \frac{1}{\pi} \left( \frac{a}{R} \right)^3 \int_0^\infty \frac{i\alpha s^3 K_1^2}{(2I_0-i\alpha s I_1)K_1+2K_0 I_1} \, \Intd s \, . \label{eq:DeltaMu_zz}
    \end{equation}
    Notably, the correction at lowest order follows a cubic dependence in the ratio of particle to cylinder radius.
    Particularly, in the hard cylinder limit we get
    \begin{equation}
    \lim_{\alpha\to\infty} \frac{\Delta \mu_{\parallel}^{\mathrm{S}}}{\mu_0^{rr}} = -\frac{1}{\pi} \left( \frac{a}{R} \right)^3 \int_0^\infty \frac{s^2 K_1}{I_1} \, \Intd s \approx 
    -0.79682 \left( \frac{a}{R} \right)^3 \, . \label{eq:DeltaMu_zz_hardCylinder}
    \end{equation}
    in agreement with the result know in the literature~\cite{brenner64, zheng92, wang95, llnton95}.
    We further emphasize that in the absence of shearing, the correction to the particle rotational mobility vanishes.

    We now turn our attention to hydrodynamic interactions between two spherical particles of equal radius~\cite{crocker97, dufresne00} positioned on the centerline of an elastic cylinder.
    For the rest of our discussion, we shall denote by $\gamma$ the particle located at $z=0$ and by $\lambda$ the particle at $z=h$.
    The particle rotational pair-mobility function about the line connecting the two centers is computed at leading order as
    \begin{equation}
    \mu_{\parallel}^{\mathrm{P}} = L_z^{-1}  \lim_{\R\to \R_\lambda} {{\Omega_1}_z} \, . \label{Delta_mu_zz_pair_def}
    \end{equation}
    Using Eqs.~\eqref{azimuthalVelocity_Axial} and \eqref{v_phi_star_Axial}, we get
    \begin{equation}
    \mu_{\parallel}^{\mathrm{P}} = \frac{1}{8\pi\eta h^3} + \frac{1}{8\pi^2\eta} \int_0^\infty q A^* \cos \left( \sigma s \right) \, \Intd q \, . \notag
    \end{equation}
    wherein $\sigma := h/R$. The first term in the equation above is the leading-order rotational pair-mobility for two isolated spheres, i.e.\@ in an unbounded medium~\cite{felderhof77}.
    Scaling by the bulk rotational mobility, we obtain
    \begin{equation}
    \frac{\mu_{\parallel}^{\mathrm{P}}}{\mu_0^{rr}} =  \left( \frac{a}{h} \right)^3  + \frac{1}{\pi} \left( \frac{a}{R} \right)^3 \int_0^\infty \frac{i\alpha s^3 K_1^2 \, \cos \left( \sigma s \right)}{(2I_0-i\alpha s I_1)K_1+2K_0 I_1} \, \Intd s  \, , \notag
    \end{equation}
    which is dependent on membrane shearing properties only.
    The hard-cylinder limit is recovered by taking $\alpha\to\infty$ to obtain
    \begin{equation}
    \lim_{\alpha\to\infty} \frac{\mu_{\parallel}^{\mathrm{P}}}{\mu_0^{rr}} =   \left( \frac{a}{h} \right)^3 -\frac{1}{\pi}  \left( \frac{a}{R} \right)^3 \int_0^\infty \frac{s^2 K_1}{I_1} \cos \left( \sigma s \right)  \, \Intd s  \, , \label{eq:DeltaMu_zz_hardCylinder_Pair}
    \end{equation}
    that is positively defined for all values of $\sigma$.
    Therefore, the two particle have always the same sense of rotation around the cylinder axis, in the same way as in an unbounded flow.

    \subsection{Radial rotational mobility}

    We compute the particle self-mobility correction for the asymmetric rotation around an axis perpendicular to the cylinder axis which for a point particle situated on the cylinder axis is
    \begin{subequations}
     \begin{align}
    \Delta \mu_{\perp}^{\mathrm{S}} = L_r^{-1} \lim_{\R\to 0} {\Omega_r^*} = L_\phi^{-1}  \lim_{\R\to 0} {\Omega_\phi^*} \, , \label{Delta_mu_rr_def} 
    \end{align}
    \end{subequations}
    where $L_r = L_x \cos \phi$ and $L_\phi = -L_x \sin \phi$ and 
    \begin{equation}
    {\Omega_r^*}    = \frac{1}{2} \left( \frac{v_{z,\phi}^*}{r} - v_{\phi, z}^* \right) \, , \qquad 
    {\Omega_\phi^*} = \frac{1}{2} (v_{r,z}^* - v_{z, r}^*) \, , \notag 
    \end{equation}
    are the corrections to the radial and azimuthal fluid angular velocity, respectively. 
    By making use of Eqs.~\eqref{v_phi_image} and \eqref{v_z_image} we get
    \begin{equation}
    \frac{\Delta \mu_{\perp}^{\mathrm{S}}}{\mu_0^{rr}} = \frac{1}{2\pi} \left( \frac{a}{R} \right)^3 \int_0^\infty \left( \gamma^* + 2{\varphi}^* \right) s^2 \, \Intd s \, . \notag 
    \end{equation}

    Considering a membrane with both shearing and bending, and by taking the vanishing frequency limit we obtain
    \begin{equation}
    \begin{split}
      \lim_{\alpha\to\infty} \frac{\Delta \mu_{\perp}^{\mathrm{S}}}{\mu_0^{rr}} &= -\frac{1}{2\pi} \left( \frac{a}{R} \right)^3
    \int_0^\infty \frac{w}{W} \, \Intd s \approx -0.73555  \left( \frac{a}{R} \right)^3 \, , \label{vanishingFreq_radial}
    \end{split}
    \end{equation}
    in agreement with the literature~\cite{brenner64, zheng92}.
    Moreover,
    \begin{subequations}
    \begin{align}
    w &=  s^2 \big( 2I_0I_1(sK_0-3K_1) + sI_0^2(sK_0+3K_1) - I_1^2 \left((s^2+8)K_0+sK_1 \right) \big) \, , \\
    W &= sI_1^3-(s^2+2)I_0I_1^2-sI_0^2I_1+s^2 I_0^3 \, .
    \end{align} \label{wAndW}
    \end{subequations}
    The same limit is obtained when considering a membrane with pure shearing.
    Another limit is recovered if the membrane possesses only a resistance towards bending such that
    \begin{equation}
    \begin{split}
      \lim_{\alpha\to\infty} \frac{\Delta \mu_{\perp, \mathrm{B}}^{\mathrm{S}}}{\mu_0^{rr}} &= -\frac{1}{\pi} \left( \frac{a}{R} \right)^3
    \int_0^\infty \frac{w_\mathrm{B}}{W_\mathrm{B}} \, \Intd s \approx -0.24688  \left( \frac{a}{R} \right)^3 \, , \label{vanishingFreq_radial_bending}
    \end{split}
    \end{equation}
    where we have defined
    \begin{align}
    w_\mathrm{B} &= s^4 K_0^2 \, , \notag \\
    W_\mathrm{B} &= s I_0 \left( (3+s^2)K_1 + 2sK_0 \right) -(3+s^2) (sK_0+2K_1) I_1 \, . \notag
    \end{align}

    Next, we turn our attention to the rotational pair-mobility perpendicular to the line of centers.
    At leading order we have
    \begin{equation}
      \mu_{\perp}^{\mathrm{P}} = L_r^{-1} \lim_{\R\to \R_\lambda} {{\Omega_1}_r} 
                               = L_\phi^{-1} \lim_{\R\to \R_\lambda} {{\Omega_1}_\phi} \, .
    \end{equation}
    In a scaled form we obtain
    \begin{equation}
    \frac{\mu_{\perp}^{\mathrm{P}}}{\mu_0^{rr}} =   -\frac{1}{2}\left( \frac{a}{h} \right)^3 + \frac{1}{2\pi} \left( \frac{a}{R} \right)^3 \int_0^\infty \left( \gamma^* + 2{\varphi}^* \right) s^2 \cos \left( \sigma s \right) \, \Intd s   \, , 
    \end{equation}
    which in the vanishing frequency limit  reduces to
    \begin{equation}
    \begin{split}
      \lim_{\alpha\to\infty} \frac{\mu_{\perp}^{\mathrm{P}}}{\mu_0^{rr}} &= -  \frac{1}{2} \left( \frac{a}{h} \right)^3 - \frac{1}{2\pi} \left( \frac{a}{R} \right)^3 
    \int_0^\infty \frac{w}{W} \, \cos \left( \sigma s \right) \, \Intd s  \, , 
    \end{split}
    \end{equation}
    with $w$ and $W$ given above by Eqs~\eqref{wAndW}. 
    {It can be shown that upon integration, the second term in the latter equation is negatively valued for all value of $\sigma$.}
    Therefore, the two particles undergo rotation in opposite directions for all values of $\sigma$, i.e. in the same way as in a bulk fluid.

    \subsection{Startup rotational motion}

    We now compute the mobility coefficients for a particle starting from rest and then rotating under a constant external torque (e.g.\@ a magnetic or optical torque) exerted either in the direction parallel or perpendicular to the cylinder axis.
    The steady torque is mathematically modeled by a Heaviside step function $\vect{L}(t) = \vect{A} \, \theta(t)$ whose Fourier transform in the frequency domain reads~\cite{bracewell99}
    \begin{equation}
    \vect{L} (\omega) = \left( \pi \delta(\omega) - \frac{i}{\omega} \right) \vect{A} \, .
    \end{equation}
    The components of the time-dependent angular velocity can readily be obtained upon inverse Fourier transformation to obtain
    \begin{equation}
     \frac{\omega_{k} (t)}{\mu_0^{rr} A_{k}} = 1 + \frac{\Delta \mu_{kk}^{\mathrm{S}}(0)}{2} 
     + \frac{1}{2i\pi} \int_{-\infty}^{+\infty} \frac{\Delta \mu_{kk}^{\mathrm{S}}(\omega)}{\omega} \, e^{i\omega t} \, \Intd \omega \, , \qquad k \in \{r,\phi,z\} \, .  \label{mobilitySteadyTorque}
    \end{equation}
    We note that the third term in Eq.~\eqref{mobilitySteadyTorque} is a real quantity which takes values between $-\Delta \mu_{kk}^{\mathrm{S}}(0)/2$ when $t\to 0$ and $+\Delta \mu_{kk}^{\mathrm{S}}(0)/2$ as $t \to \infty$, corresponding to the bulk and hard-wall behaviors, respectively.
    As the frequency-dependent mobilities are expressed as integrals over the scaled wavenumber $s$, the computation of the time-dependent angular velocities requires a double integration procedure.
    For this aim, we use the Cuba Divonne algorithm~\cite{hahn05, hahn16} which is found to be suitable for the numerical computation of the present double integrals.

    \subsection{Membrane deformation}

    Finally, our results can be employed to compute the membrane deformation resulting from an arbitrary time-dependent point-torque acting parallel or perpendicular to the cylinder axis. 
    The membrane displacement field can readily be computed from the fluid velocity at $r=R$ via the non-slip relation stated by Eq.~\eqref{no-slip-relation}. 
    We define the membrane frequency-dependent reaction tensor in the same way as previously defined for a point-force as~\cite{bickel07,daddi16b}
    \begin{equation}
    u_{i} (z,\phi, \omega) = Q_{{ij}} (z,\phi, \omega) L_{j} (\omega)\, , \label{VerformungVonDerKraft}
    \end{equation}
    relating between the membrane displacement field $\vect{u}$ and the torque $\vect{L}$ acting on the nearby particle.
    Considering a harmonic-type driving torque $L_i (t) = A_i e^{i\omega_0 t}$, the membrane deformation in the time domain is calculated as 
    \begin{equation}
    u_{i} (z,\phi, t) = Q_{{ij}} (z,\phi,\omega_0) A_{j} e^{i\omega_0 t} \, .
    \end{equation}
    The physical displacement is then obtained by taking the real part of the latter equation. 
    From Eqs.~\eqref{v_phi_2_Axial} and \eqref{A_2}, we obtain
    \begin{equation}
    Q_{\phi z}  =  \Lambda \int_0^\infty  \frac{2s K_1 (I_0K_1 + I_1 K_0)}{(2K_0-i\alpha s K_1)I_1 + 2I_0 K_1} \cos \left( \frac{sz}{R} \right) \, \Intd s \, , \notag 
    \end{equation}
    wherein $\Lambda := {1}/{(4 i \pi^2 \eta \omega R^2)}$,  giving access to the membrane azimuthal deformation when an axial torque is exerted on the particle.
    We further have $Q_{rz} = Q_{zz} = 0$ due to symmetry.

    Next, considering a torque acing along an axis perpendicular to the cylinder axis, we obtain 
    \begin{align}
    Q_{r\phi}  &=  - \Lambda \int_0^\infty \bigg( \left( (2 + s^2)K_1+s K_0 \right){\varphi}_2
		-\left(sK_0+K_1\right)\psi^* - K_1 \, \gamma^* \bigg) \sin \left( \frac{sz}{R} \right) \, \Intd s \, , \notag \\
    Q_{\phi r} &= \Lambda  \int_0^\infty \bigg( -\left(s K_0+2K_1  \right){\varphi}_2 
		+ K_1\psi^* + \left( K_1+sK_0 \right)\gamma^* \bigg) \sin \left( \frac{sz}{R} \right) \, \Intd s \, , \notag \\
    Q_{z\phi}  &=  -\Lambda \int_0^\infty s \left( -s K_0{\varphi}_2 + K_1 \psi^* \right) \cos \left( \frac{sz}{R} \right) \, \Intd s  \, ,  \notag 
    \end{align}
    and $Q_{rr} = Q_{\phi\phi} = Q_{zr} = 0$.

    \section{Comparison with numerical simulations}\label{comparisonWithBIM}

    \begin{figure}
    \begin{center}
      \includegraphics[scale=0.95]{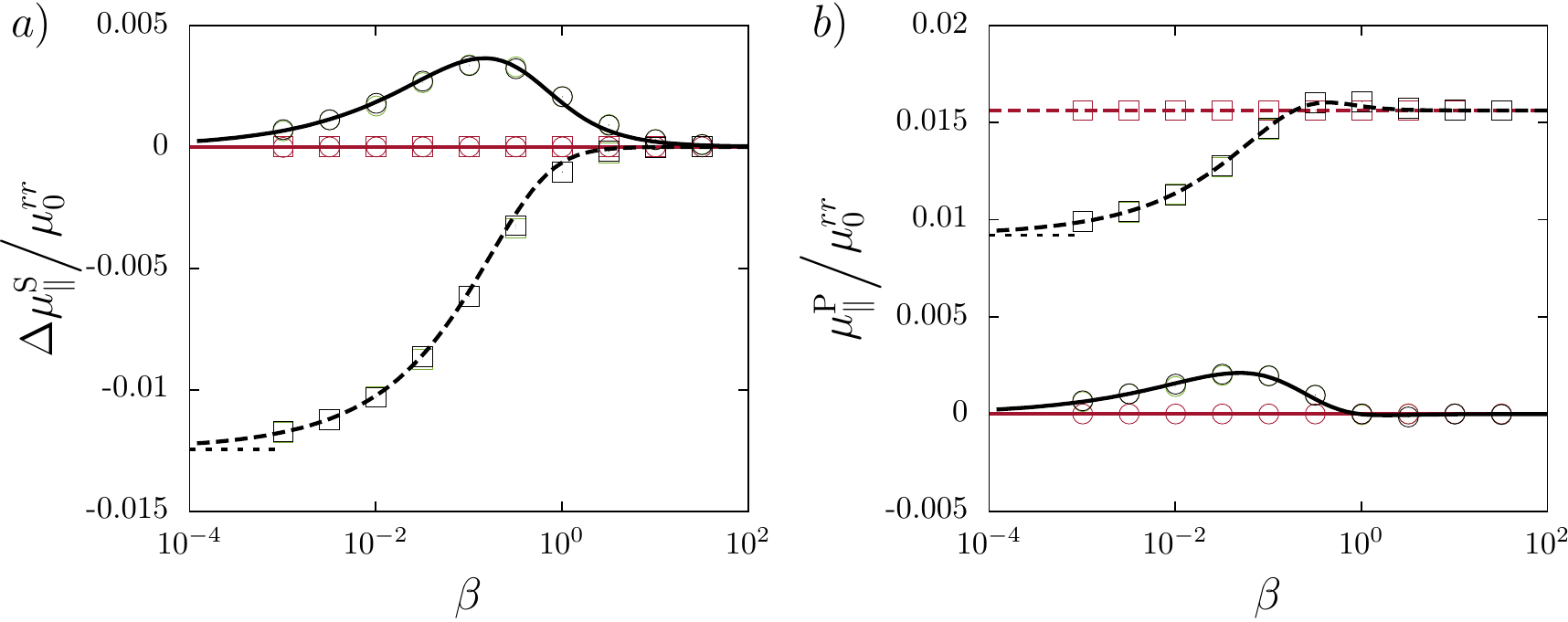}
    \caption{(Color online) The scaled frequency dependent self- $(a)$ and pair- $(b)$ mobilities versus the scaled frequency $\beta$ for the rotational motion around the cylinder axis. The membrane is endowed with only-shearing (green), only-bending (red) or both shearing and bending rigidities (black). Green lines/symbols are hardly visible as they overlap with the black lines/symbols. Here the particle is set on the centerline of an elastic cylindrical membrane of radius $R=4a$. For the pair-mobility, the interparticle distance is set $h=R$.
    The analytical predictions are shown as dashed and solid lines for the real and imaginary parts, respectively. BIM simulations are presented as squares and circles for the real and imaginary parts, respectively. The horizontal dashed lines represent the hard-cylinder limits predicted by Eq.~\eqref{eq:DeltaMu_zz_hardCylinder} and \eqref{eq:DeltaMu_zz_hardCylinder_Pair} for the self- and pair-mobilities respectively. For other parameters, see main text.
    } 
    \label{cylindricoRad_RR_ZZ}
    \end{center}
    \end{figure}

    \begin{figure}
    \begin{center}
      \includegraphics[scale=0.95]{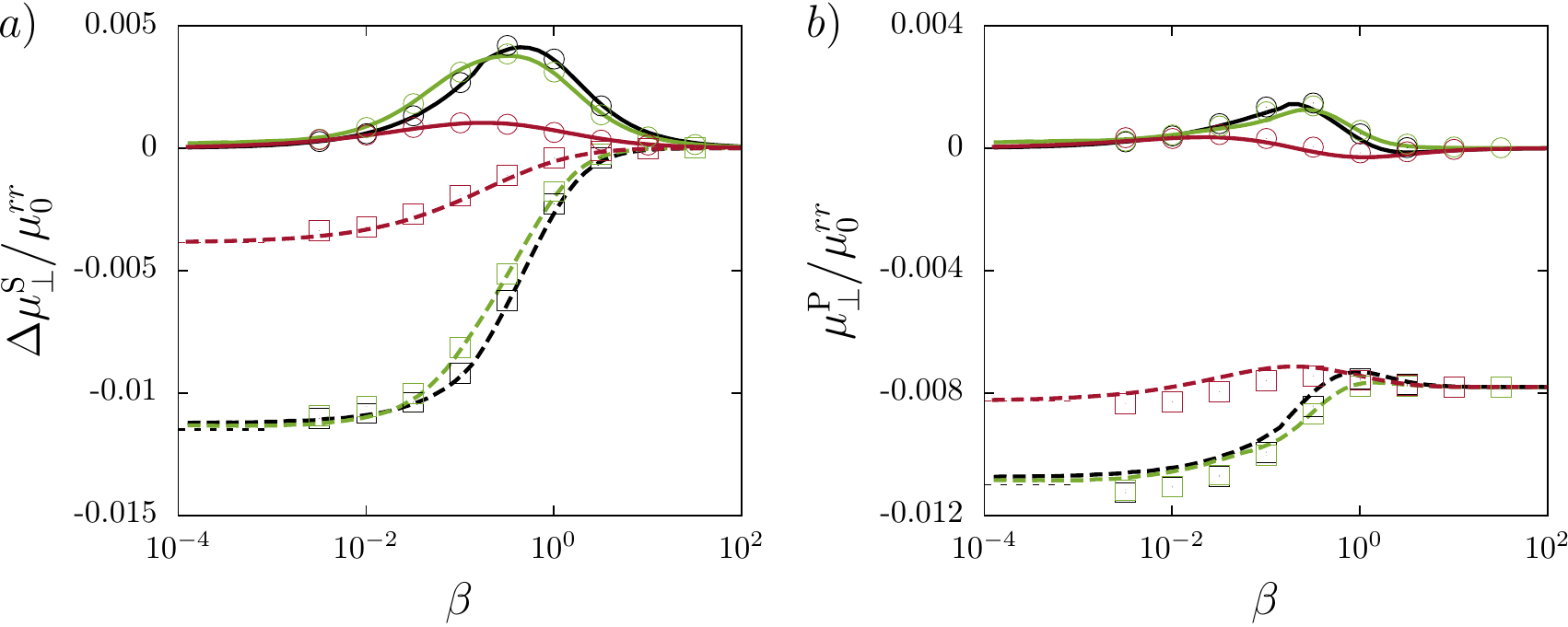}
    \caption{(Color online) The scaled frequency dependent self- $(a)$ and pair- $(b)$ mobilities versus the scaled frequency $\beta$ for the rotational motion around an axis perpendicular to the cylinder axis. The color code is the same as in Fig.~\ref{cylindricoRad_RR_ZZ}.
    } 
    \label{cylindricoRad_RR_XX}
    \end{center}
    \end{figure}

    \begin{figure}
    \begin{center}
      \includegraphics[scale=1.2]{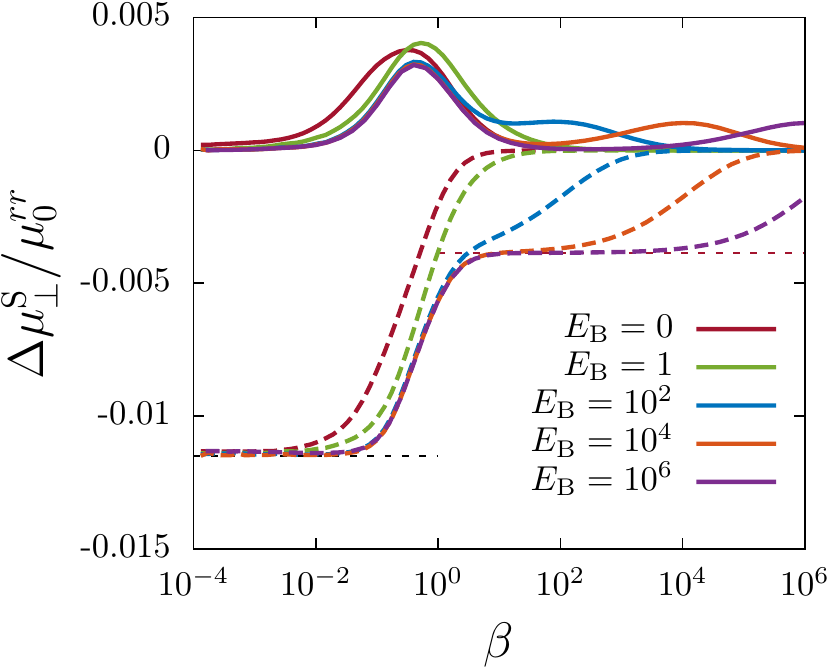}
    \caption{{(Color online) Scaled particle self-mobility corrections versus $\beta$ for various values of the reduced bending modulus $\EB$ for the rotational motion around an axis perpendicular to the cylinder axis. Here we take $R=4a$ and $C=1$.}} 
    \label{cylindricoRad_EB_TT_XX}
    \end{center}
    \end{figure}

     \begin{figure}
    \begin{center}
      \includegraphics[scale=0.95]{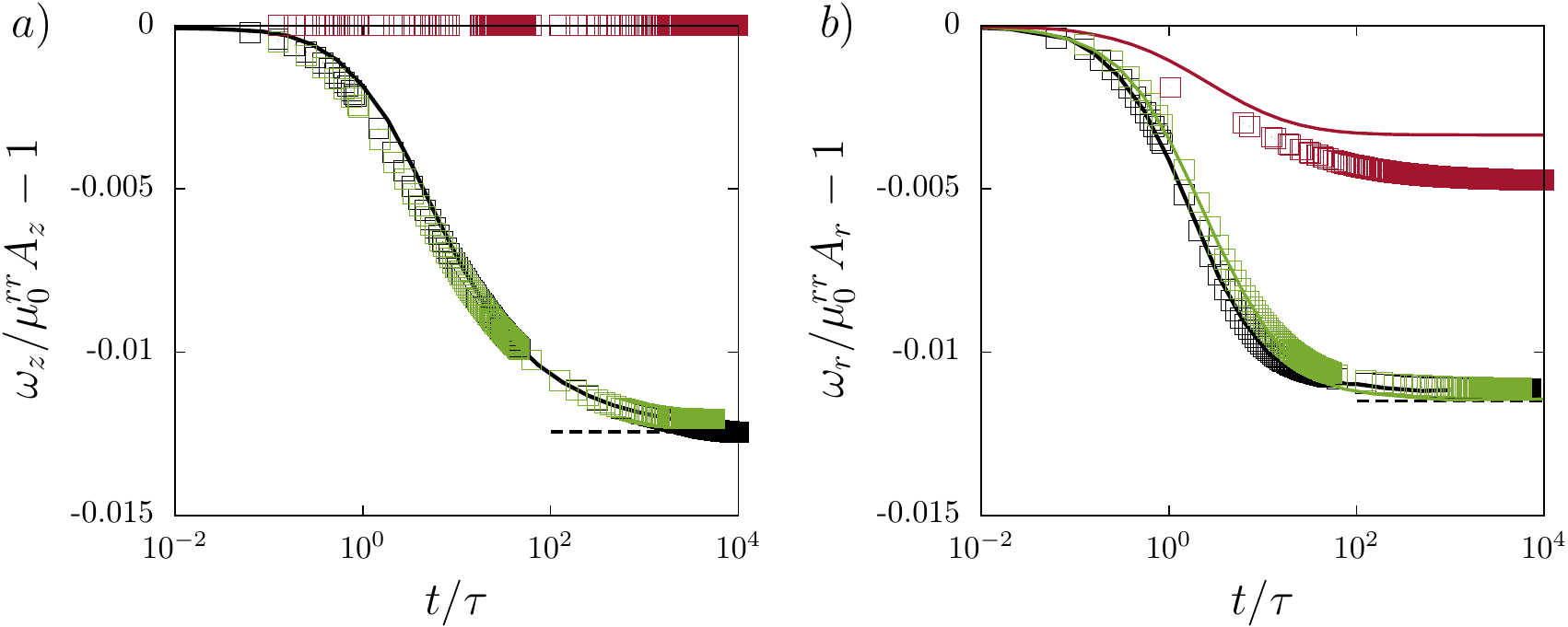}
    \caption{(Color online) Time-dependent angular velocity of a particle starting from rest for $a)$ axial and $b)$ radial rotational motion under the action of a constant external torque.
    Here we use the same parameters as in Fig.~\ref{cylindricoRad_RR_ZZ} with a membrane with both shearing and bending rigidities. 
    Solid lines are the analytical predictions obtained from Eq.~\eqref{mobilitySteadyTorque} whereas symbols are the boundary integral simulations results. Black dashed lines are our theoretical predictions based on the point particle approximation. Here $\tau$ is a characteristic time scale defined as $\tau := \beta/\omega$.}  
    \label{steadyMotion_Torque}
    \end{center}
    \end{figure}

    In order to assess the validity and appropriateness of the point particle approximation employed throughout this work, we compare our analytical predictions with computer simulations performed using a completed double layer boundary integral equation method~\cite{phan93, phan94, kohr04, zhao11, zhao12}.
    The method is known to be ideally suited for simulation of fluid flows in the Stokes regime~\cite{pozrikidis01} where both solid objects and deformed boundaries are present.
    Technical details regarding the method and its numerical implementation have been reported by some of us elsewhere, see e.g. Refs.~~\cite{daddi16b, guckenberger16}.

    In the simulations, the cylindrical membrane is of length $200a$ uniformly meshed with 6550 triangles.
    The spherical particle is discretized by 320 triangular elements obtained by refining an icosahedron consecutively~\cite{krueger11, kruger16book}.

    In order to compute numerically the particle rotational self- and pair-mobility functions, a time dependent harmonic torque ${L_\lambda}_{i} (t) = {A_\lambda}_{i} e^{i\omega_0 t}$ of amplitude ${A_\lambda}_{i}$ and frequency $\omega_0$ is exerted along the direction ${i}$ at the particle labeled $\lambda$ either parallel ($z$ direction) or perpendicular ($x$ direction) to the cylinder axis.
    After a short transient evolution, both particles undergo oscillatory rotation with the same frequency $\omega_0$ but with different phases, such that ${\Omega_\lambda}_{i} = {B_\lambda}_{i} e^{i\omega_0 t + \delta_\lambda}$ and ${\Omega_\gamma}_{i} = {B_\gamma}_{i} e^{i\omega_0 t + \delta_\gamma}$.
    For an accurate determination of the angular velocity amplitudes and phase shifts, we use a nonlinear least-squares solver based on the trust region method~\cite{conn00}.
    The particle rotational self- and pair-mobilities are then computed as
    \begin{equation}
    \mu_{{ij}}^{\mathrm{S}} = \frac{{B_\lambda}_{i}}{{A_\lambda}_{j}} \, e^{i\delta_\lambda} \, , \quad\quad  
    \mu_{{ij}}^{\mathrm{P}} = \frac{{B_\gamma}_{i}}{{A_\lambda}_{j}} \, e^{i\delta_\gamma} \, .
    \end{equation}

    We then define the characteristic frequency associated to shearing as $\beta := 1/\alpha = 3\eta\omega R/(2\kS)$, and for bending as $\betaB := 1/\alphaB^3 = \eta\omega R^3/\kB$.
    Additionally, we define the reduced bending modulus $\EB := \kB/(\kS R^2)$ a parameter quantifying the relative effect between membrane shearing and bending.

    As an example setup, we place a spherical particle on the centerline of an elastic cylinder of initial (undeformed) radius $R=4a$.
    We mostly take a reduced bending modulus $\EB = 1/6$ for which the characteristic frequencies associated to shearing and bending have about the same order of magnitude, and thus both effects manifest themselves equally.

    Fig.~\ref{cylindricoRad_RR_ZZ}~$a)$  shows the parallel component of the correction to the rotational self-mobility function upon variation of the forcing frequency $\beta$.
    For a membrane with bending-only resistance (shown in red), both the real and imaginary parts of the mobility correction vanish, in agreement with our theoretical prediction stated by Eq.~\eqref{eq:DeltaMu_zz}.
    Not surprisingly, the torque exerted on the particle along the cylinder axis induces only membrane torsion and therefore the resulting stresses do not cause any  out-of-plane deformation or bending.
    For a membrane with a non-vanishing shearing resistance, however, we observe that the mobility correction exhibits a monotonically increasing real part and the typical peak structure for the imaginary part. 
    In the vanishing frequency limit, the correction to rotational mobility is identical to that predicted nearby a hard-cylinder with stick boundary conditions, given by Eq.~\eqref{eq:DeltaMu_zz_hardCylinder}.
    Moreover, the bulk behavior is recovered for large forcing frequencies where both the real and imaginary parts vanish.

    In Fig.~\ref{cylindricoRad_RR_ZZ}~$b)$ we present the rotational pair-mobility function for two particles located on the cylinder centerline a distance $h=R$ apart.
    Similarly, a membrane with pure bending resistance does not introduce a correction to the particle pair-mobility.
    Yet, the latter is markedly affected by the membrane shearing resistance where the correction approaches that near a hard-cylinder in the low frequency regime.
    For high forcing frequencies, the pair mobility equals that of two equal-sized spheres in an unbounded medium, given at leading order by $(a/h)^3$.
    A good agreement is obtained between theoretical predictions and the numerical simulations we performed using a completed double layer boundary integral method.

    \begin{figure}
    \begin{center}
      \includegraphics[scale=1]{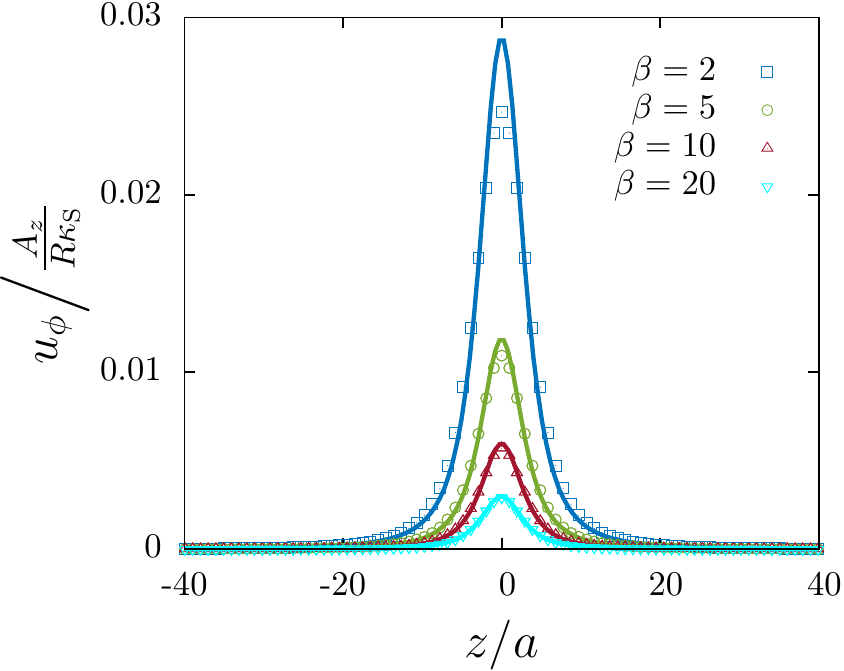}
    \caption{(Color online) The scaled azimuthal membrane displacements versus $z/a$ at four forcing frequencies computed at quarter oscillation period for $t\omega_0 = \pi/2$. Solid lines are the analytical predictions and symbols refer to boundary integral simulations.} 
    \label{deformation_Rotational_ZZ}
    \end{center}
    \end{figure}

    \begin{figure}
    \begin{center}
      \includegraphics[scale=1]{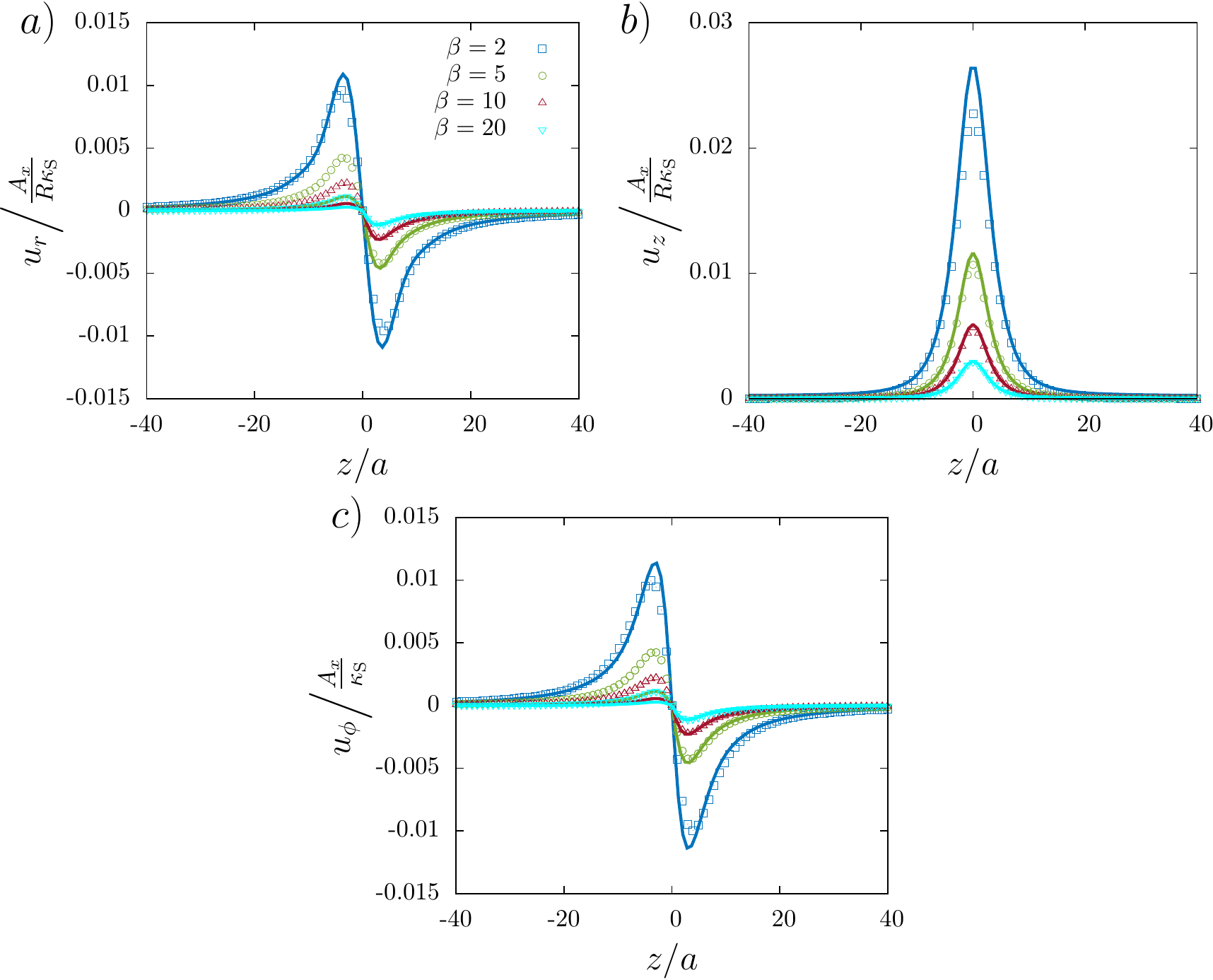}
    \caption{(Color online) The scaled radial $a)$, axial $b)$ and azimuthal $c)$ membrane displacement field versus scaled distance along the axis $z/a$ for four forcing frequencies calculated at quarter period for $\omega_0 t=\pi/2$ when the particle reaches its maximal radial position. 
    Here deformations are shown in the plane of maximum deformation.
    Solid lines are the theoretical predictions and symbols refer to the boundary integral simulation results. } 
    \label{deformation_Rotational_XX}
    \end{center}
    \end{figure}

    We now carry out for the rotation about an axis perpendicular to the cylinder axis.
    In Fig.~\ref{cylindricoRad_RR_XX}, we show the perpendicular component of the particle rotational self- and pair-mobilities nearby  a membrane endowed with shearing-only (green), bending-only (red) or both shearing and bending rigidities (black).
    The mobility functions show basically a similar evolution as in the previous case of axisymmetric rotation around the cylinder axis. 
    As explained before, we observe that the mobility near a no-slip cylinder is recovered only if the membrane possesses a non-vanishing shearing resistance.
    The pair-mobility in the high frequency regime can appropriately be estimated from the leading order bulk pair-mobility $-(1/2)(a/h)^3$.

    {In order to probe the effect of the aforementioned coupling between shear and bending, we show in Fig.~\ref{cylindricoRad_EB_TT_XX} the particle self-mobility function versus $\beta$ for the rotational motion perpendicular to the cylinder axis upon variation of the reduced bending modulus $\EB$ while keeping $R=4a$.
    We observe that as $\EB$ increases, a second peak of lower amplitude emerges for higher forcing frequencies in the imaginary part.
    Additionally, a dispersion step in the real part occurs that bridges between the hard-cylinder limit Eq.~\eqref{vanishingFreq_radial} and the bending limit predicted by Eq.~\eqref{vanishingFreq_radial_bending}.
    In fact, the peak observed at $\beta\sim 1$ is attributed to the membrane resistance towards shear and can conveniently be estimated by a simple balance between fluid viscosity and membrane elasticity as $\omega \sim \kS/(\eta R)$.
    The high-frequency peak is however attributed to the membrane resistance towards bending and its position can properly be estimated by a balance between fluid viscosity and bending such that $\omega \sim \kB/(\eta R^3)$ corresponding to  $\betaB \sim 1$.
    Since $\betaB = 2\beta/(3\EB)$, the second peak occurs at $\beta \sim \EB$.
    Particularly, for $\EB=1$, the shear and bending related peaks coincide for which both effects manifest themselves equally.
    Analogous predictions can be made for the pair-mobility where similar conclusions can be drawn.
    }

    In Fig.~\ref{steadyMotion_Torque}, we show the time-dependent angular velocity of a particle initially at rest, rotating under the action of a constant external torque.
    We scale the time by the characteristic time scale for shearing defined as $\tau := \beta/\omega = 3\eta R/(2\kS)$.
    At short time scales for which $t \ll \tau$, the membrane introduces a small correction to the particle mobility since it does not have enough time to react on these short time scales.
    As the time increases, the membrane effect becomes more important and the mobility curves bend down substantially to asymptotically approach the correction predicted nearby a hard cylinder.
    The steady rotational mobilities undergo small corrections relative to the bulk values, making them more difficult to obtain precisely from the simulations. This explains the small discrepancy between theory and simulations, notably for a membrane with pure bending resistance.

    The membrane displacement induced by the symmetric rotation of the particle around the cylinder axis is shown in Fig.~\ref{deformation_Rotational_ZZ} where both analytical predictions (solid lines) and numerical simulations (symbols) are presented.
    Here we use the same parameters as in Fig.~\ref{cylindricoRad_RR_ZZ} and four different actuation frequencies.
    Displacement fields are plotted when the oscillating particle reaches its maximal angular position.
    We observe that the membrane azimuthal deformation exhibits a bell-shaped behavior that peaks at the origin where deformation is more pronounced. 
    By comparing the membrane displacement field at various forcing frequencies, we observe that as the forcing frequency gets larger, the membrane undergoes remarkably smaller deformation since the membrane does not have sufficient time to respond to the fast rotating particle.

    Analogous predictions for asymmetric deformation induced by the particle radial rotation are shown in Fig.~\ref{deformation_Rotational_XX}. 
    Here deformations are shown in the plan of maximum deformation, i.e. $\phi=0$ for $u_\phi$, and $\phi=\pi/2$ for $u_r$ and $u_z$.
    The radial and azimuthal deformations have fundamentally the same evolution where both have symmetry with respect to the origin at which the deformation vanishes. 
    On the other hand, axial deformation reaches its maximal value at the origin and decays far away as the ratio $z/a$ gets larger.
    It can clearly be seen that upon particle radial rotation, the membrane undergoes primarily axial deformation with a maximum that is about three times larger than that reached in the radial or azimuthal deformations. 

    For typical flow parameters, the torques exerted by optical tweezers on suspended nanoparticles are of the order of 1~pN~$\mu$m~\cite{friese01}.
    Assuming a cylinder radius of $10^{-6}~$m, a membrane shearing modulus of about $ 10^{-6}~$N/m  and an actuation frequency $\beta = 2$, the membrane undergoes a maximal deformation of about 3~\% of its undeformed radius.
    Therefore, deformations upon particle rotational motion are small and deviations from cylindrical shape are negligible.

    \section{Conclusions}\label{conclusions}

    In this contribution we have presented analytical calculations of the Stokes flow induced by a point-torque exerted parallel or perpendicular to the axis of an elastic circular tube.
    The membrane is modeled by a combination of the neo-Hookean model for shearing and Helfrich model for bending.
    The solution of the fluid flow is expressed in terms of Fourier-Bessel integrals with unknown coefficients which are determined from the boundary conditions imposed at the membrane.

	The result is the Green's function for two orientations of the rotlet singularity. In the limit when shearing and bending coefficients are large, corresponding to a stiff membrane, our results converge to the expressions previously derived in literature for a hard cylindrical no-slip tube. 
    
    Our results are directly applicable to the determination of the leading-order correction to the self- and pair-mobility functions of particles rotating parallel or perpendicular to the cylinder axis. Notably, the correction to self mobility follows a cubic dependence on the ratio of particle to cylinder radius. We also find that the rotational mobilities along the axis depend solely on membrane shearing resistance and that bending does not play any role. Both shearing and bending however manifest themselves for the rotational motion along an axis perpendicular to the cylinder axis.  More importantly, the steady particle mobility nearby a hard-cylinder with stick boundary conditions is recovered only if the membrane possesses a non-vanishing resistance towards shearing. As an example, we have calculated the effects of startup motion, i.e. particle initially at rest starting to rotate under a steady torque.
        The Green's function can also be applied to the calculation of the resulting membrane deformation. For realistic values of parameters, however, this turns out to be negligible. 
   
    Our analytical predictions are verified and supplemented by corresponding boundary integral simulations where a good agreement is obtained.


    \begin{acknowledgements}
    ADMI and SG thank the Volkswagen Foundation for financial support and acknowledge the Gauss Center for Supercomputing e.V. for providing computing time on the GCS Supercomputer SuperMUC at Leibniz Supercomputing Center. 
    Additionally, they thank Achim Guckenberger for fruitful discussions.
    This work has been supported by the Ministry of Science and Higher Education of Poland via the Mobility Plus Fellowship awarded to ML.
    This article is based upon work from COST Action MP1305, supported by COST (European Cooperation in Science and Technology).
    \end{acknowledgements}

    \section{Membrane mechanics} \label{appendix:membraneMech}

    In this Appendix, the traction jump across a membrane endowed with shearing and bending rigidities will be derived in the cylindrical coordinates system.
    We denote by $\vect{a} = R \eR + z \eZ$ the position vector of the points located at the undeformed membrane.
    Here $R$ is the membrane (undeformed) radius and $z$ is the axial distance along the cylinder axis.
    Here $r$, $\phi$ and $z$ refer to the radial, azimuthal and vertical coordinates, respectively.
    After deformation, the vector position reads
    \begin{equation}
    \vect{r} = (R + u_r)\eR + u_\phi \ePhi + (z + u_z) \eZ \, ,
    \end{equation}
    where $\vect{u}$ denotes the displacement vector field, which depends on the in-plane variables $\phi$ and $z$.
    In the following, we shall use capital Roman letters for the undeformed state and small roman letters for the deformed.
    The cylindrical membrane is defined by the covariant base vectors $\gOne := \vect{r}_{,\phi}$ and $\gTwo := \vect{r}_{,z}$, which read
    \begin{align}
    \gOne &=  (u_{r,\phi} - u_\phi)\eR + (R + u_r + u_{\phi, \phi}) \ePhi  + u_{z,\phi} \eZ \, , \\
    \gTwo &=  u_{r,z} \eR + u_{\phi, z} \ePhi + (1 + u_{z,z} ) \eZ \, .
    \end{align}
    The unit normal vector $\vect{n}$ is defined as
    \begin{equation}
    \vect{n} = \frac{\gOne \times \gTwo}{|\gOne \times \gTwo|}  \, ,
    \end{equation}
    which, at leading order in deformation reads
    \begin{equation}
    \vect{n} = \eR + \frac{u_\phi - u_{r, \phi}}{R} \ePhi - u_{r,z} \eZ \, . 
    \end{equation}

    The covariant components of the first fundamental form (metric tensor) are defined by the scalar product $g_{{ij}} = \vect{g}_{{i}} \cdot \vect{g}_{{j}}$.
    Upon linearization, we obtain 
    \begin{equation}
    g_{{ij}} = \left(
		      \begin{array}{cc}
		      R^2 + 2R (u_r + u_{\phi, \phi}) & u_{z,\phi} + R u_{\phi, z} \\
		      u_{z,\phi} + R u_{\phi, z}  & 1+2u_{z,z} 
		      \end{array}
		      \right) \, .  \label{covariantTensor}
    \end{equation}

    The contravariant tensor $g^{{ij}}$ is the inverse of the metric tensor~\cite{deserno15}, and at leading order reads
    \begin{equation}
    g^{{ij}} = \left(
		      \begin{array}{cc}
		      \frac{1}{R^2} - 2\frac{u_r + u_{\phi, \phi}}{R^3} & -\frac{u_{z,\phi} + R u_{\phi, z}}{R^2} \\
		      -\frac{u_{z,\phi} + R u_{\phi, z}}{R^2}  & 1-2u_{z,z} 
		      \end{array}
		      \right) \, .
			\label{contravariantTensor}
    \end{equation}

    The covariant and contravariant tensors in the undeformed state $G_{{ij}}$ and $G^{{ij}}$ can immediately be obtained by considering a vanishing displacement in Eqs.~\eqref{covariantTensor} and \eqref{contravariantTensor}, respectively.
    In the following, the traction jump equations across a cylindrical membrane endowed by an in-plane shearing resistance shall be derived.

    \subsection{Shearing}

    The two transformation invariants are given by Green and Adkins as~\cite{green60, zhu14}
    \begin{subequations}
    \begin{align}
      I_1 &= G^{{ij}}  g_{{ij}}  -  2 \, , \\
      I_2 &= \det G^{{ij}}  \det g_{{ij}}  -  1 \, .
    \end{align}
    \end{subequations}

    The contravariant components of the stress tensor $\tau^{{ij}}$ can readily be obtained from the membrane constitutive relation such that~\cite{lac04}
    \begin{equation}
    \tau^{{ij}} = \frac{2}{\JS} \frac{\partial W}{\partial I_1} G^{{ij}} + 2\JS \frac{\partial W}{\partial I_2} g^{{ij}} \, ,
    \label{stressTensor}
    \end{equation}
    where $W(I_1, I_2)$ is the areal strain energy density and $\JS := \sqrt{1+I_2}$ is the Jacobian determinant, representing the  ratio  between  the  deformed  and undeformed local surface area.
    In the linear theory of elasticity, $\JS \simeq 1 + e$, where $e := (u_r+u_{\phi,\phi})/R + u_{z,z}$ is the dilatation function.
    In the present work, we use the neo-Hookean model to describe the elastic properties of the membrane, whose areal strain energy reads~\cite{krueger12, zhu15}
    \begin{equation}
    W(I_1, I_2) = \frac{\kS}{6} \left( I_1 - 1 + \frac{1}{1+I_2} \right) \, .
    \label{skalakEquation}
    \end{equation}

    Plugging Eq.~\eqref{skalakEquation} into Eq.~\eqref{stressTensor}, the linearized in-plane stress tensor reads
    \begin{equation}
    \tau^{{ij}} = \frac{2 \kS}{3}
    \left(
    \begin{array}{cc}
      \frac{u_r + u_{\phi, \phi}}{R^3} + \frac{e}{R^2} & \frac{1}{2R} \left( u_{\phi, z} + \frac{u_{z, \phi}}{R} \right) \\
      \frac{1}{2R} \left( u_{\phi, z} + \frac{u_{z, \phi}}{R} \right) & u_{z,z} + e 
    \end{array}
    \right) \, .
    \end{equation}

    The membrane elastic forces are balanced by the external forces via the equilibrium equations
    \begin{subequations}
    \begin{align}
      \nabla_{{i}} \tau^{{ij}} + \Delta f^{{j}} &= 0 \, , \label{Equilibrium_Tangential} \\
      \tau^{{ij}} b_{{ij}} + \Delta f^{n} &= 0 \, , \label{Equilibrium_Normal}
    \end{align}
    \label{Equilibrium}
    \end{subequations}
    where $\Delta \vect{f} = \Delta f^{{j}} \vect{g}_{{j}} + \Delta f^{n} \vect{n} $ is the traction jump vector across the membrane.
    Here $\nabla_{{i}}$ stands for the covariant derivative~\cite{synge69} and $b_{{ij}}$ is the second fundamental form (curvature tensor) defined by the dot product $b_{{ij}} = \vect{g}_{{i},{j}} \cdot \vect{n}$.
    At leading order we obtain
    \begin{equation}
    b_{{ij}} = 
    \left(
    \begin{array}{cc}
    u_{r,\phi\phi} - (R + u_r + 2 u_{\phi, \phi}) & u_{r, \phi z} - u_{\phi, z} \\
    u_{r, \tehta z} - u_{\phi, z} & u_{r,zz}
    \end{array}
    \right) \, . \label{curvatureTensor}
    \end{equation}

    After some algebra, the traction jump equations across the membrane given by Eqs.~\eqref{Equilibrium} read
    \begin{subequations}
    \begin{align}
    \frac{\kS}{3} \left( u_{\phi, zz} + \frac{3 u_{z, \phi z}}{R} + \frac{4 (u_{r,\phi} + u_{\phi, \phi\phi})}{R^2} \right) + \Delta f_\phi &=0 \, , \\
      \frac{\kS}{3} \left( 4 u_{z,zz} + \frac{2 u_{r,z} + 3 u_{\phi, z\phi}}{R} + \frac{u_{z, \phi\phi}}{R^2} \right) + \Delta f_{z} &= 0 \, , \\
      -\frac{2\kS}{3} \left( \frac{2 (u_r + u_{\phi, \phi}) }{R^2} + \frac{u_{z,z}}{R}  \right) + \Delta f_{r} &= 0 \, .
    \end{align}
    \label{Equilibrium_finalize}
    \end{subequations}

    Continuing, the jump in the fluid stress tensor across the membrane reads
    \begin{equation}
    [ \sigma_{{j} r} ] = \Delta f_{{j}} \, , \quad {j} \in \{ z, r \} \, . \label{stressTensorTractionJump}
    \end{equation}

    Therefore, From Eqs.~\eqref{Equilibrium_finalize}, \eqref{stressTensorTractionJump} and \eqref{no-slip-relation}, it follows that 
    \begin{subequations}
      \begin{align}
	[v_{\phi,r} ] &= \left. \frac{i\alpha}{2} \left( R v_{\phi, zz} + {3 v_{z, \phi z}} + \frac{4 (v_{r,\phi} + v_{\phi, \phi\phi})}{R} \right) \right|_{r=R} \, , \\
	[v_{z,r} ] &= \left. \frac{i\alpha}{2} \left( 4 R v_{z,zz} + {2 v_{r,z} + 3 v_{\phi, z\phi}} + \frac{v_{z, \phi\phi}}{R} \right) \right|_{r=R} \, , \\
	\left[-\frac{p}{\eta} \right] &= \left. -i\alpha \left( \frac{2 (v_r + v_{\phi, \phi} )}{R} + v_{z,z} \right) \right|_{r=R} \, , 
      \end{align}
    \end{subequations}
    where $\alpha := 2\kS/(3 \eta R \omega) $ is the shearing coefficient.
    Note that it follows from the incompressibility equation 
    \begin{equation}
    \frac{v_r + v_{\phi, \phi}}{r} + v_{r,r} + v_{z,z} = 0 \, , 
    \end{equation}
    that $[v_{r,r}] = 0$.
    In the following, we shall derive the traction jump equations across a membrane with pure bending rigidity.


    \subsection{Bending}

    We use the Helfrich model, in which the traction jump equations across the membranes are given by~\cite{daddi16, Guckenberger_2017}
    \begin{equation}
    \Delta \vect{f}  = -2\kB \left( 2(H^2-K+H_0 H) + \Delta_\parallel \right) (H-H_0) \, \vect{n} \, ,  
    \end{equation}
    where $\kB$ is the bending modulus, $H$ and $K$ are respectively the mean and Gaussian curvatures, given by
    \begin{equation}
    H = \frac{1}{2} \, b_{i}^{i} \, , \qquad 
    K = \mathrm{det~} b_{i}^{j} \, , 
    \end{equation}
    with $b_{i}^{j}$ being the mixed version of the curvature tensor related to the covariant representation of the curvature tensor by $b_{{i}}^{{j}} = b_{{i}{k}} g^{{k}{j}}$.
    Continuing, $\Delta_\parallel$ is the Laplace-Beltrami operator and $H_0$ is the spontaneous curvature, for which we take the initial undisturbed shape here. 
    The linearized traction jump due to bending are therefore given by 
    \begin{equation}
    \begin{split}
      -\kB \Big( & R^3 u_{r,zzzz} + 2R(u_{r,zz} + u_{r,zz\phi\phi})  
	+ \frac{u_r+2u_{r,\phi\phi}+u_{r, \phi\phi\phi\phi}}{R} \Big)  + \Delta f_r = 0 \, .
    \end{split}
    \end{equation}
    and $\Delta f_\phi = \Delta f_z = 0$.

    Note that bending does not introduce at leading order a jump in the tangential traction~\cite{Guckenberger_2017}.
    The traction jump equations take the following final from
    \begin{subequations}
    \begin{align}
      [v_{\phi, r}] &= 0 \, , \\ 
      [v_{z,r} ] &= 0 \, , \\
      \left[-\frac{p}{\eta} \right] &= - i\alphaB^3 \Big( R^3 v_{r,zzzz} + 2R(v_{r,zz} + v_{r,zz\phi\phi}) 
	   + \left. \frac{v_r+2v_{r,\phi\phi}+v_{r, \phi\phi\phi\phi}}{R} \Big) \right|_{r=R} \, , 
    \end{align}
    \end{subequations}
    where $\alphaB = (\kB /(\eta \omega))^{1/3}/R$ is the bending coefficient.



\end{document}